\documentclass[11pt,a4paper]{article}
\usepackage[margin=2.4cm]{geometry}
\usepackage{amsmath,amssymb}
\usepackage{physics}
\usepackage{booktabs}
\usepackage{microtype}
\usepackage{graphicx}
\usepackage{array}
\usepackage[colorlinks=true,linkcolor=blue,citecolor=blue,urlcolor=blue]{hyperref}
\usepackage{mathtools}
\usepackage[capitalise]{cleveref}

\newcommand{\Leff}{\Lambda_{\mathrm{eff}}}
\newcommand{\Lcore}{\Lambda_{\mathrm{core}}}
\newcommand{\rhoMS}{\rho_{\mathrm{MS}}}
\newcommand{\weff}{w_{\mathrm{eff}}}
\newcommand{\rH}{r_{H}}
\newcommand{\Hinf}{H_{\mathrm{inf}}}
\newcommand{\Ltil}{\widetilde{\Lambda}}

\title{\textbf{Regular Cosmologies:\\ Three Distinct Routes}}
\author{Sergio Bravo Medina$^{1}$\thanks{sergiobravom@javeriana.edu.co. ORCID: 0000-0003-4399-1281} \and Marek Nowakowski$^{2,3}$\thanks{marek.nowakowski@ictp-saifr.org. ORCID: 0000-0001-8716-4670}\\[4pt]
\small $^{1}$Grupo de F\'isica Te\'orica (GFT), Departamento de F\'isica,
Pontificia Universidad Javeriana,\\
\small Cra.\ 7 No.\ 40-62, Bogot\'a, Colombia\\
\small $^{2}$ICTP--South American Institute for Fundamental Research,
S\~ao Paulo, SP, Brazil\\
\small $^{3}$Departamento de F\'isica, Universidade Federal de S\~ao Paulo
(UNIFESP), Diadema, SP, Brazil}
\date{\today}

\begin{document}
\maketitle

\begin{abstract}
\noindent
Regular
black holes replace the central singularity with a de~Sitter core. We ask what cosmology follows
when the same core is carried to the expanding universe, through the position-dependent
cosmological term $\Leff(r)=8\pi G\rhoMS(r)$ of six regular metrics. There is no unique way to do
this, and we follow three routes: a Newtonian one after McCrea and Milne, a running vacuum
evaluated at the Hubble radius, and a quasi-local one based on the Misner--Sharp mass. The
Newtonian route gives a universe that starts in a de~Sitter phase and ends as dust, solved in closed form for the Hayward metric. We show that it is the flat FLRW sector of the
four-dimensional quasi-topological theory whose vacuum is the regular black hole, so that for the
metrics that descend from such a theory the inflationary core is derived and not postulated. The
Hubble-radius routes are phenomenological deformations, in which the core contributes a bounded vacuum term that ordinary matter dominates toward the past. In all three routes a monotonically decreasing core density never
crosses $w=-1$, so the phantom behaviour suggested by DESI would require a density shell that a
regular core does not have. At late times the routes differ. We compare them indicatively
with the DESI DR2 results. A single regularisation length cannot serve as both the inflationary
and the dark-energy scale.

\end{abstract}

\noindent\textit{Keywords}: regular black holes, de Sitter core, non-singular cosmology,
quasi-topological gravity, running vacuum, dark energy


\section{Introduction: three routes from one idea}
\label{sec:intro}

The singularity theorems \cite{Penrose65,Hawking70} guarantee that, within general relativity and under
standard energy conditions, gravitational collapse ends in a  singularity. Regular black holes are the best-studied way of evading this
conclusion at the level of the metric: one replaces the Schwarzschild interior
with a non-singular core. This core is almost always a patch of de Sitter space, in which the
curvature is finite and the strong energy condition is violated. Some regular solutions like Bardeen's model~\cite{Bardeen1968}, Hayward's
model~\cite{Hayward2006}, and Dymnikova's vacuum nonsingular black
hole~\cite{Dymnikova1992,Dymnikova2002} share this feature: written as
$f(r)=1-2Gm(r)/r$, their mass function grows as $m(r)\sim r^{3}$ near the centre,
so the metric behaves as $f\to1-(\Lcore/3)r^{2}$: de Sitter, with a finite
$\Lcore$.

The metrics that achieve this
are, for the most part, educated guesswork: one posits a geometry with the
desired regularity and asymptotics and identifies a source only afterwards, as in
the nonlinear-electrodynamics reading of Ayón-Beato and García~\cite{AyonBeato1998,AyonBeato2000}. They are
taken seriously nonetheless, because a singularity-free spacetime is expected to
be a low-energy shadow of the correct theory of quantum gravity, and these models
are our best guess at its shape. The same idea can be brought to early
cosmology and that is what we attempt here. It may be that the removal of the
big-bang singularity and of the black-hole singularity rests on common ground, in
the way that a single quantum principle like the existence of a bounded-below
ground state of the vacuum makes the
atom stable against classical collapse, produces the Casimir force between
conductors, and regularises the black-body spectrum by curing the ultraviolet
catastrophe. The idea that a bounded curvature should replace both singularities is
older than most of the regular metrics we use: it is Markov's limiting-curvature
hypothesis~\cite{Markov1982}, realised for black-hole interiors as a de~Sitter core by
Frolov, Markov and Mukhanov~\cite{FMM1989} and for cosmology by Mukhanov, Brandenberger and Sornborger~\cite{MB1992,BMS1993}, whose nonsingular universe asymptotes to de~Sitter
in the past exactly as one of the routes (route N) will. The present work can be read as asking which of
the explicit regular metrics realise that hypothesis cosmologically, and under which map.

Dymnikova observed long ago that the de Sitter core can be read as a
\emph{variable cosmological term} \cite{Dymnikova2002} defining
\begin{equation}
  \Leff(r)\equiv 8\pi G\,\rhoMS(r),\qquad
  \rhoMS(r)=\frac{m'(r)}{4\pi r^{2}},
  \label{eq:Leffdef}
\end{equation}
where $\rhoMS$ is the (coordinate-invariant) Misner--Sharp energy
density \cite{MisnerSharp1964}, one has a function that saturates to
$\Lcore$ at the centre and decays to zero at spatial infinity. The object
\eqref{eq:Leffdef} is the bridge between black-hole interiors and cosmology: it
is a vacuum energy density that is large where the geometry is most curved and
negligible in the asymptotic region, a behaviour one would
want of a term that drove early acceleration but gets switched off later.

To turn this analogy into a cosmology one must make two choices that are
logically independent: a \emph{map} relating the metric radius $r$ to a
cosmological scale, and a \emph{prescription} fixing what density and pressure
the metric supplies to the Friedmann system. Given that different reasonable choices exist,  we organise the paper around three of them, which
turn out to populate different epochs and expose different physical behaviours:

\begin{itemize}\setlength\itemsep{3pt}
\item \emph{Route N (Newtonian / McCrea--Milne).} The oldest and most
  conservative construction. One treats a homogeneous expanding sphere of the
  regular-black-hole ``fluid'' and reads off the Friedmann equation from
  Newtonian energy balance~\cite{Milne1934,McCreaMilne1934,FaraoniAtieh2020}.
  Here the regular black hole is the \emph{sole} source, the comoving radius is
  identified with the scale factor, $R\propto a$. The construction
 is a controlled heuristic that reproduces the relativistic
Friedmann pair: the energy equation directly, and the acceleration equation,
complete with its $(\rho+3p)$ source, once the first law is used to eliminate
$\dot\rho$ (\cref{sec:routeN}). What remains truly Newtonian is that pressure never
sources the potential.
\item \emph{Route A/B (running vacuum).} One promotes
  $\Leff(r)$ to a function of the scale factor through the Hubble-radius
  identification $r\rightarrow \rH=H^{-1}$ and treats the result as a running vacuum
  in the sense of Shapiro and Sol\`a~\cite{ShapiroSola2002} and Sol\`a~\cite{Sola2013}, added on
  top of ordinary matter and radiation. We will show that the radial equation of
  state of the core, $p_{r}=-\rhoMS$, is what justifies this reading.
\item \emph{Route C (quasi-local / Faraoni).} One uses the
  Misner--Sharp--Hernandez mass enclosed within the cosmological apparent
  horizon~\cite{HernandezMisner1966,Faraoni2015}, $R_{\rm AH}=H^{-1}$ for a
  spatially flat universe, as the gravitating energy that sources the expansion.
\end{itemize}

The three are not competitors but rather complement each other: route N is most
informative about the early universe behaviour and singularity resolution, routes A/B and
C are informative about the late-time expansion and its confrontation with data. They agree
  on the core value the geometry supplies, differ on whether that value is
reached (\cref{ssec:coresurvival}), and differ again where the late-time prescription matters.
For the metrics that descend from a quasi-topological theory, route N is moreover the FLRW
sector of that theory (\cref{sec:covariant}), which settles its status. We aim to track that agreement and disagreement in this work.

Route N is singularity-free for every de Sitter-core metric and fully analytic for Hayward. Its cosmography shows that the acceleration it produces is
\emph{primordial}, an inflationary core with a graceful exit to a dust era, not a
late-time dark sector (\cref{sec:routeN,sec:routeNsol,sec:routeNcosmo}).  An argument that runs route by route (\cref{sec:criteria}) shows that no
\emph{monotone} core can cross the phantom divide in any of them, so the dynamical dark energy
favoured by DESI requires a density shell, which among our metrics only
Jusufi--Singleton develops and only in its self-energy-dominated regime, where its authors themselves regard the unbounded self-energy as unphysical and cap it (\cref{sec:criteria}). The
running-vacuum route A/B reaches the DESI $w_{0}$ band but fails to match its $|w_{a}|$
(\cref{sec:routeAB,sec:routeABcosmo}), and the quasi-local route C sits
closer to $\Lambda$CDM (\cref{sec:routeC,sec:routeCimpl}). We close with a classification of the different routes with respect to the models and an outlook (\cref{sec:classification}).

The idea that a regular black-hole interior carries a possible cosmology has been
explored and improved considerably in recent work, and it is useful to state
how the present paper compares to these publications. In the work by Neves \cite{Neves2017} the mass function
is taken into the scale factor of an inhomogeneous metric, obtaining a bounce but losing
the cosmological principle. The constructions relevant here keep homogeneity and map
the metric into a Friedmann equation. A Swiss-cheese matching of Hayward, Bardeen and Dymnikova black holes to an expanding
universe, with an accelerated phase as its outcome, is studied in \cite{Dialektopoulos2025}, and the dark-energy component induced by such a matching is confronted with observations in \cite{Abilmazhinova2026}. The interior trapped region of a static
regular black hole is a Kantowski--Sachs instead of an FLRW spacetime (homogeneous but with spherical rather than flat spatial sections, it expands differently along and across the radial direction),
this is made explicit by Casadio, Kamenshchik and
Ovalle~\cite{CasadioKO2025}, who construct regular Schwarzschild black holes
parametrised by the ADM mass alone and analyse their (anisotropic,
quasi-cyclic) Kantowski--Sachs interior cosmologies. The minimal way to obtain
an \emph{isotropic} daughter is instead a junction matching, developed in the
Oppenheimer--Snyder spirit by Li, Wu and Ge~\cite{LiWuGe2025} for
Minkowski-core metrics, and analysed in general by
Easson~\cite{Easson2026}. In that sense the static black hole is identified as the parent
and the spatially flat FLRW region glued to it is the daughter: so the daughter is not postulated but
inherited, its Friedmann equation being forced by the Darmois junction conditions from the parent's mass
function, $H^{2}+k/a^{2}=2Gm(a\chi_{b})/(a\chi_{b})^{3}$, which is the
relativistic counterpart of the Newtonian relation we use in route N
(\cref{sec:routeN}). The matching is minimal: a no-shell matching whose
evolution is fixed by the parent alone, with no added late-time bulk sector, no modified asymptotics
and no independent shell stress tensor. Easson
then proves an obstruction: such a minimal,
no-shell, asymptotically flat daughter cannot be simultaneously indefinitely
expanding, curvature-regular, geodesically complete and consistent with the
averaged null energy condition (non-negative $T_{\mu\nu}k^{\mu}k^{\nu}$ integrated along a complete null geodesic). The closed branch is bounded by the finite
ADM mass: the flat and open branches are past null incomplete, meaning
that a past-directed null geodesic runs out of spacetime at finite affine parameter
$\lambda=\int a dt=\int da/H$, even where the span of cosmic time is infinite. 
Of the three exclusions in the definition of minimality, routes A/B and C will
later violate only the first. This
obstruction tells us what a
\emph{minimal} construction cannot do, and the escape it identifies, ``an
additional smooth bulk component whose density redshifts no faster than
$A^{-2}$,'' $A$ being the scale factor there, of which a vacuum-energy component is the simplest example, is precisely what
routes A/B and C supply. The present work
is therefore complementary on three fronts: where the minimal daughter is the
endpoint of those analyses, it is our \emph{starting} point (route N), whose
analytic solutions, critical-exponent classification and cosmography we work out in full across six metrics. The running-vacuum and quasi-local routes are the
``additional structure'' the obstruction calls for and the crossing criterion of
\cref{sec:criteria}, on the equation of state rather than on global completeness, is
a statement none of these works make. A separate strand on the hydrodynamic
stability of the anisotropic source with a prescribed equation of
state \cite{Firouzjahi2026}, matches part of our analysis through the
radial relation $p_{r}=-\rho$ and the role of the weak energy condition, to
which we return in \cref{sec:criteria,sec:routeAB}.

Cosmology has been approached from a different direction. In
quasi-topological gravity (QTG) with infinite towers of higher-curvature terms regular
black holes arise as the unique vacuum solutions in $D\ge5$ \cite{BuenoCano2025}, with
Dymnikova type metrics obtained from the mechanism in \cite{KonoplyaZhidenko2024}, and the
corrections that make the black hole regular do the same for its cosmology: Oppenheimer--Snyder
collapse becomes a bounce \cite{BCHMV2025}, FLRW singularities are eliminated in the non-polynomial
$d=4$ versions \cite{BorissovaMagueijo2026}, and a brane moving in a regular black hole bulk inflates, with a de Sitter
phase set by the regularisation scale and a universal e-fold count $N\simeq \frac{D-3}{D-1}\ln(r_g/\sqrt{\alpha})$
\cite{Sueto2026}. These regularisations remain in the early universe and derive from higher dimensions
or specific towers while the present paper starts from given four dimensional metrics, maps them
by three routes and carries them to late-time data. Both approaches meet on the inflationary core, which in route N
yields an M-independent de Sitter phase with logarithmic e-fold count (\cref{sec:routeNcosmo}).

\section{Regular black holes and the choice of metrics}
\label{sec:metrics}

We take the static, spherically symmetric line element
\begin{equation}
  \dd s^{2}=-f(r)\,\dd t^{2}+f(r)^{-1}\dd r^{2}+r^{2}\dd\Omega^{2},
  \qquad f(r)=1-\frac{2Gm(r)}{r},
  \label{eq:ssmetric}
\end{equation}
with mass function $m(r)$. Regularity at $r=0$ requires $m(r)\to\tfrac{1}{2G}c\,r^{3}$
for some constant $c$, so that $f\to1-c\,r^{2}$ and the curvature invariants are
finite. The constant then fixes the core value $\Lcore=3c$ of the effective term
\eqref{eq:Leffdef}. In the list below we set $G=1$ and restore it where the dependence matters. We study six metrics, chosen to span the physically distinct
mechanisms by which regularity (or its absence) is realised, being at the
same time the most-used representatives in the literature. These metrics are:

\begin{itemize}\setlength\itemsep{2pt}
\item \textbf{Hayward}~\cite{Hayward2006}: $f=1-2Mr^{2}/(r^{3}+2M\ell^{2})$, the
  minimal rational model with a single length $\ell$, the reference case.
\item \textbf{Bardeen}~\cite{Bardeen1968,AyonBeato2000}:
  $f=1-2Mr^{2}/(r^{2}+g^{2})^{3/2}$, the original regular solution, sourced by a
  magnetic monopole in nonlinear electrodynamics.
\item \textbf{Dymnikova}~\cite{Dymnikova1992}: $f=1-(2M/r)(1-e^{-r^{3}/r_{0}^{3}})$,
  the vacuum nonsingular black hole with a Gaussian-like density.
\item \textbf{Nicolini--Smailagic--Spallucci (NSS)}~\cite{NSS2006}: a
  noncommutative-geometry-inspired metric whose point mass is smeared into a
  Gaussian of width $\sqrt{\theta}$, with $m(r)\propto\gamma(\tfrac32,r^{2}/4\theta)$ (in the tables and figures we set $\theta=\ell^{2}/2$).
\item \textbf{Jusufi--Singleton (JS)}~\cite{JusufiSingleton2025}: a recent
  neutral metric sourced by gravitational self-energy, motivated as a
  dark-matter candidate. 
Its mass function is the T-duality (Bardeen) core of the smeared source plus
  the regularised energy of its own Newtonian field,
  \begin{equation}
  m(r)=\frac{M r^{3}}{(r^{2}+\ell^{2})^{3/2}}
  +\frac{3M^{2}}{16\ell}\arctan\frac{r}{\ell}
  -\frac{M^{2}r\,(5r^{2}+3\ell^{2})}{16\,(r^{2}+\ell^{2})^{2}},
  \label{eq:mJS}
  \end{equation}
  so that its Misner--Sharp density is Bardeen-class with a self-energy bump,
  $\rhoMS=3M\ell^{2}/4\pi(r^{2}+\ell^{2})^{5/2}+M^{2}r^{2}/8\pi(r^{2}+\ell^{2})^{3}$,
  and its ADM mass is $\mathcal{M}=m(\infty)=M(1+3\pi M/32\ell)$, larger than the bare
  $M$ by the field energy. (\Cref{eq:mJS} is written with $G=1$. Restoring it, the two self-energy terms and the correction to $\mathcal{M}$ carry one more factor of $G$.) We will see that the bump is the only place in our gallery
  where a density shell can form.
\item \textbf{Bonanno--Reuter}~\cite{BonannoReuter2000}: the
  renormalisation-group-improved (asymptotically safe) metric, in which the
  Newton coupling runs $G(r)=G_{0}r^{3}/[r^{3}+\tilde\omega G_{0}(r+\gamma G_{0}M)]$.
  This is the member of the set whose regularisation comes from a
  renormalisation-group flow rather than a chosen profile. We use the values $\gamma=9/2$ and
  $\tilde{\omega}=118/15\pi$ of \cite{BonannoReuter2000} and let the overall scale of 
  $G_{0}$ float so that the core value is $\Lcore=6/(\gamma\tilde{\omega}G_{0})\equiv 3/\ell^{2}$, where $\ell$ is then the single regularisation length quoted in the tables. The fiducial mass is fixed through $G_{0}M=\ell/2$, the same ratio of gravitational radius to core length as $M=\ell/2$ for the other metrics.
   (Having an origin beyond a chosen profile no longer singles this metric out: covariant action principles reproducing the
  Hayward metric have recently been exhibited, from non-polynomial pure
  gravity~\cite{BorissovaCarballoRubio2026} and from vector-tensor
  theories~\cite{EichhornFernandes2026}.)
\end{itemize}

For the homogeneous map we keep the regular black hole neutral, since a static
electric charge has no net counterpart in an isotropic universe. The
Ayón-Beato-Garc\'ia electric metric is taken up separately in
\cref{ssec:rung}. We take this case as a two-scale deformation rather than a literal
cosmological charge, where decoupling the charge scale from the regularisation
length leaves one component with a fixed equation of state, which in $H(z)$ scales as $(1+z)^{5}$.

\begin{figure}[t]
  \centering
  \includegraphics[width=\textwidth]{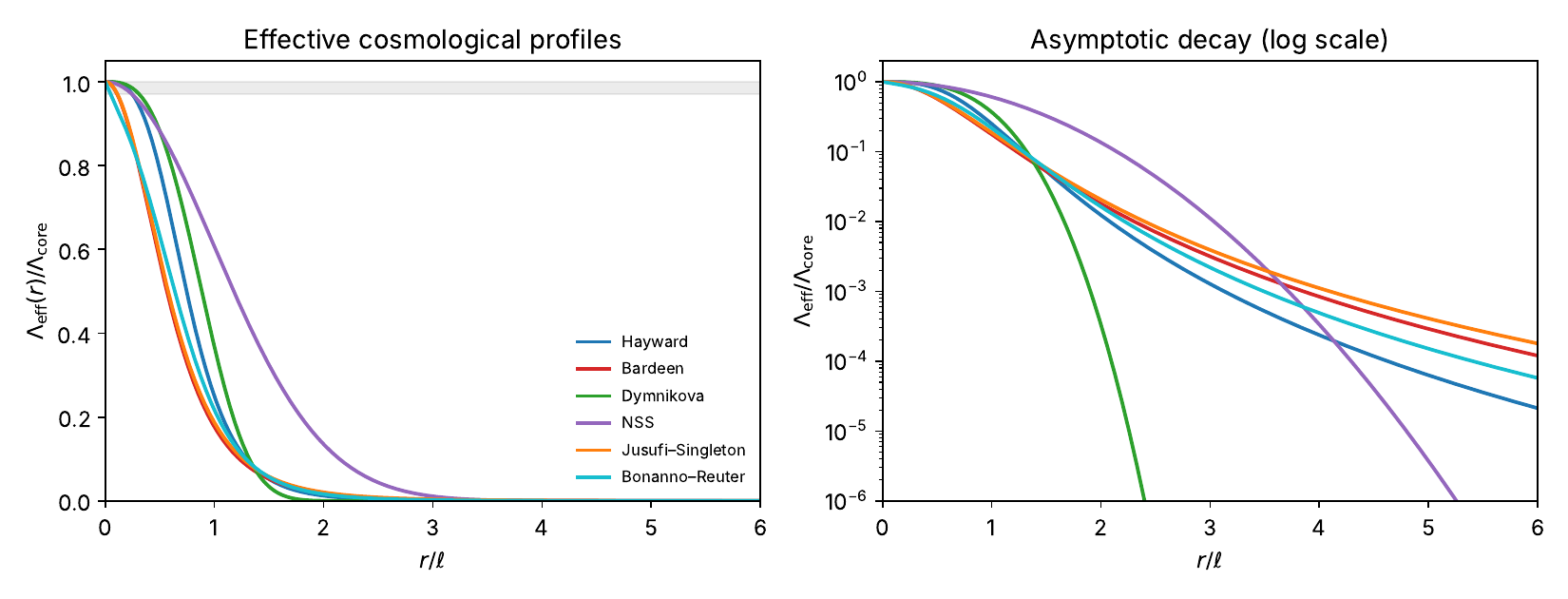}
  \caption{The effective cosmological term $\Leff(r)/\Lcore$ for the six
    metrics, on linear (left) and logarithmic (right) vertical axes
    ($M=\ell/2$, with $\ell$ denoting each metric's own regularisation scale,
    in units of the core length). All saturate to the
    de~Sitter core value as $r\to0$ (shaded band) and decay to zero. The decay
    law governs the late-time equation of state but not the universal core.}
  \label{fig:profiles}
\end{figure}

The two profile functions we extract from each $f$ are the
Misner--Sharp density and the effective cosmological term,
\begin{equation}
  \rhoMS(r)=\frac{m'(r)}{4\pi r^{2}},\qquad
  \Leff(r)=8\pi G\,\rhoMS(r)=\frac{1}{r^{2}}\frac{\dd}{\dd r}\!\big[r(1-f)\big].
  \label{eq:profiles}
\end{equation}
\Cref{fig:profiles} shows $\Leff(r)/\Lcore$ for all six: each saturates to the
de~Sitter plateau at small $r$ and decays at large $r$ with a metric-dependent
decay law: rational for Hayward ($r^{-6}$), Bardeen and Bonanno--Reuter ($r^{-5}$) and
JS ($r^{-4}$, the fall-off of the Newtonian field energy carried by its self-energy
term), and faster than any power for the Gaussian Dymnikova and NSS. This feature
distinguishes the late-time cosmology while the universal core does not.

\section{Route N: Friedmann equations}
\label{sec:routeN}

The McCrea--Milne construction obtains the Friedmann equation from Newtonian
mechanics~\cite{Milne1934,McCreaMilne1934}. Consider a homogeneous sphere of
proper radius $R(t)$, density $\rho$ and mass $M_{\rm s}=\tfrac{4\pi}{3}R^{3}\rho$, and take
a test galaxy of mass $\mu_{\rm t}$ which on its surface obeys energy conservation,
$\tfrac12\mu_{\rm t}\dot R^{2}-GM_{\rm s}\mu_{\rm t}/R=E$. Dividing by
$\tfrac12\mu_{\rm t}R^{2}$ and using $\dot R/R=\dot a/a=H$,
\begin{equation}
  H^{2}=\frac{8\pi G}{3}\rho-\frac{k}{R^{2}},
  \qquad k\equiv-\frac{2E}{\mu_{\rm t}},
  \label{eq:MMfriedmann}
\end{equation}
which is exactly the relativistic first Friedmann equation, with the integration
constant playing the role of spatial curvature. The construction is a celebrated
case of obtaining the correct result for a partly wrong reason: the acceleration
equation that follows from Newton's second law, $\ddot a/a=-\tfrac{4\pi G}{3}\rho$,
omits the relativistic pressure term, because in Newtonian gravity pressure does
not gravitate~\cite{FaraoniAtieh2020}. The omission, however, belongs to the force law and not to the construction as
a whole. The energy equation (\ref{eq:MMfriedmann}), together with the first law
applied to the comoving volume, $\dot\rho+3H(\rho+p)=0$, already determines
the acceleration equation: differentiating the former and eliminating
$\dot\rho$ through the latter returns
$\ddot a/a=-\tfrac{4\pi G}{3}(\rho+3p)$ exactly, and the same manipulation
goes through unchanged for every $\ell\neq0$ profile of \cref{sec:routeNsol}. Newton's
second law is then recognised as the dust special case of this pair: the
energy balance with $m=\tfrac{4\pi}{3}R^{3}\rho$ presupposes $\dot m=0$,
whereas for $p\neq0$ the first law reads
$\dot m=-p\,\tfrac{d}{dt}\big(\tfrac{4\pi}{3}R^{3}\big)$, and imposing the
force law on top of the energy equation and continuity over-determines the
system. What route N cannot do is let pressure source the potential $\Phi$. The pressure sector it carries is the one dictated by conservation,
which is also the one the matched FLRW region carries through the Bianchi
identity. We therefore treat route N as a generator of the full Friedmann
pair, with pressure entering through the continuity instead of
through the force law.

The construction can also be used as a generator of modified cosmologies. Its
only dynamical input is the potential, so any correction to $\Phi(R)$ propagates
directly into $H^{2}$ without passing through the field equations, as was used
in Ref.~\cite{BargBMNB2016} to import the one-loop quantum corrections of the
Newtonian potential into a set of quantum-corrected Friedmann equations, whose
solutions include a bounce of the kind familiar from loop quantum cosmology.
Route N is the same machinery driven by a different input. Here, instead of a
perturbative correction, we insert the potential defined by a regular
black-hole metric function and the work done there by the quantum correction is
done here by the de~Sitter core.

The route is made cosmological by feeding it the regular-black-hole potential.
Writing the weak-field specific potential as $\Phi(R)=\tfrac{1}{2}[f(R)-1]$
and substituting $M\to\tfrac{4\pi}{3}R^{3}\rho$ (equivalently, by using
the mass function directly with $R\propto a$), \cref{eq:MMfriedmann} reads
\begin{equation}
  H^{2}(a)=\frac{1-f(a)}{a^{2}}=\frac{2G\,m(a)}{a^{3}}
          =\frac{8\pi G}{3}\,\bar\rho_{\rm MS}(a),
  \qquad \bar\rho_{\rm MS}(a)=\frac{3\,m(a)}{4\pi a^{3}},
  \label{eq:routeNgeneric}
\end{equation}
so the source is the \emph{mean} Misner--Sharp density inside the comoving
sphere. The regular black hole is the entire content of the universe, i.e.\  there is no separate matter or radiation.

This same relation has been derived in a more rigorous way and it is worthwhile to see how
both derivations are connected. Let us glue a spatially flat FLRW region to a
static regular black hole across a comoving sphere $\chi_{b}$ and demand no surface layer
at the junction (the Darmois conditions, i.e. continuity of the induced metric and of the extrinsic
curvature). The FLRW side is then forced to obey \cite{Easson2026}
\begin{equation}
  H^{2}+\frac{k}{a^{2}}=\frac{2G\,m(a\chi_{b})}{(a\chi_{b})^{3}}.
  \label{eq:junction}
\end{equation}
For $k=0$ this is exactly \cref{eq:routeNgeneric} absorbing the comoving radius into the normalisation, so the Newtonian
argument and the gluing calculation lead to the same equation. Additionally, the junction derivation makes a geometrical
point which we will adopt: the interior of a static regular black hole is not an FLRW spacetime but a Kantowski--Sachs one~\cite{CasadioKO2025,Easson2026}. Route N is therefore to be understood as a homogeneous model
which happens to share the Friedmann equation of the matched region (not as a rewriting of the black-hole interior).
We will come back to the global structure of this model and to the obstruction which constrains the matched region in \cref{ssec:coresurvival}. Each metric yields an explicit $H^2 (a)$
so that, for instance, Hayward behaves as follows:
\begin{equation}
  H^{2}_{\rm Hayward}(a)=\frac{2GM}{a^{3}+2GM\ell^{2}}
  \;\xrightarrow{a\to\infty}\;\frac{2GM}{a^{3}}
  \;\xrightarrow{a\to0}\;\frac{1}{\ell^{2}},
  \label{eq:routeNhayward}
\end{equation}
an Einstein--de~Sitter dust tail at late times and a de~Sitter plateau at early
times. The corresponding $H^{2}$ for the other metrics is collected in \cref{tab:routeNcore}. In every de~Sitter-core case the small-$a$ limit is the
finite $\Lcore/3$, while the large-$a$ limit is the dust tail $2GM/a^{3}$.

We make the small-$a$ limit explicit, because the way the
black-hole mass enters is easy to misread. The core value is a \emph{density}, and it is
mass-independent. Quite generally, the de~Sitter-core condition is that the
central Misner--Sharp density is finite, $\rhoMS(0)\equiv\rho_{\rm core}<\infty$,
so that the mass function fills the volume, namely
\begin{equation}
  m(a)\;\xrightarrow{a\to0}\;\frac{4\pi}{3}\,\rho_{\rm core}\,a^{3},
  \label{eq:coremass}
\end{equation}
and substituting into \cref{eq:routeNgeneric} the factor $a^{3}$ cancels between
numerator and denominator,
\begin{equation}
  H^{2}=\frac{2Gm(a)}{a^{3}}\;\xrightarrow{a\to0}\;\frac{8\pi G}{3}\,\rho_{\rm core}
       =\frac{\Lcore}{3}.
  \label{eq:corelimit}
\end{equation}
For Hayward one sees the cancellation directly: as $a\to0$,
$m(a)=Ma^{3}/(a^{3}+2GM\ell^{2})\to a^{3}/(2G\ell^{2})$, the mass $M$ dropping
out, so $\rho_{\rm core}=3/(8\pi G\ell^{2})$ and $H^{2}\to1/\ell^{2}$ carry no
trace of $M$. The black-hole mass therefore sets the \emph{transition} scale (the value of
$a$ at which the universe leaves the core, $a^{3}\sim2GM\ell^{2}$), not the
core value itself, which is a pure vacuum scale fixed by the
regularisation length $\ell$ alone. This is physically consistent: the
de~Sitter core is a property of how the singularity is cured, not of the mass of
the object, and the same statement holds for each metric in
\cref{tab:routeNcore} with its own core density (e.g.\ $\rho_{\rm core}=3M/4\pi g^{3}$
for Bardeen, with $H^{2}=\Lcore/3$ through the cancelled $a^{3}$).

The small-$a$ behaviour of route N admits two seemingly different descriptions:
a dynamical one, in which ``something other than matter'' drives an early
accelerated expansion, and an energetic one, in which the enclosed mass is tied
to the density and one worries that it diverges at the origin. They are the same
statement seen from two sides and writing both
explicitly dismisses the apparent danger. We separate the enclosed mass into the part fixed by the central density and the remainder. For any de~Sitter-core
metric,
\begin{equation}
  m(a)=\underbrace{\frac{4\pi}{3}\,\rho_{\rm core}\,a^{3}}_{\text{core (vacuum) part}}
       \;+\;\underbrace{\big[m(a)-\tfrac{4\pi}{3}\rho_{\rm core}a^{3}\big]}_{\text{matter-like remainder}},
  \label{eq:massdecomp}
\end{equation}
where the first term is fixed by the finite central density $\rho_{\rm core}=\rhoMS(0)$
and the second collects the higher-order fall-off of the profile. Inserting this
into $H^{2}=2Gm(a)/a^{3}$ gives
\begin{equation}
  H^{2}(a)=\frac{8\pi G}{3}\,\rho_{\rm core}
          \;+\;\frac{2G}{a^{3}}\big[m(a)-\tfrac{4\pi}{3}\rho_{\rm core}a^{3}\big],
  \label{eq:H2split}
\end{equation}
in which the first term is a constant, an effective cosmological constant
$\Lcore=8\pi G\rho_{\rm core}$, and the second is what would, for a
non-regular profile, carry a matter-like $a^{-3}$ divergence.

As $a\to0$ the remainder in \cref{eq:H2split} vanishes
(for Hayward, $m(a)-a^{3}/2G\ell^{2}=-a^{6}/[2G\ell^{2}(a^{3}+2GM\ell^{2})]\to0$),
leaving
\begin{equation}
  H^{2}\;\xrightarrow{a\to0}\;\frac{\Lcore}{3},\qquad
  a(t)\;\longrightarrow\;a_{i}\,e^{\Hinf t},\qquad
  \Hinf=\sqrt{\frac{\Lcore}{3}}=\frac{1}{\ell},
  \label{eq:inflate}
\end{equation}
an exponential, accelerated, de~Sitter phase driven by the
core vacuum energy $\rho_{\rm core}$. This is the regular-black-hole counterpart
of a Starobinsky-type early acceleration~\cite{Starobinsky1980}: the role of the
inflaton potential's plateau is played by the finite core density, and the
inflationary scale $\Hinf=1/\ell$ is set by the regularisation length instead of
a scalar sector. The deceleration parameter there is $q=-1$,
maximal acceleration, as for any constant-$H$ phase. The full cosmography of
the transition is explored in \cref{ssec:cosmography}.

A different way to view the same limit answers the worry that the enclosed mass
``explodes at the singularity.'' Indeed, the opposite happens. Because the central density
is finite, the mass inside a vanishing comoving sphere \emph{vanishes},
\begin{equation}
  m(a)\;\xrightarrow{a\to0}\;\frac{4\pi}{3}\,\rho_{\rm core}\,a^{3}\;\to\;0,
  \label{eq:massvanish}
\end{equation}
and this $a^{3}$ is what cancels the $a^{3}$ in the denominator of
$H^{2}=2Gm/a^{3}$ to leave the finite plateau \eqref{eq:inflate}. A Schwarzschild
core would have $m\to M\ne0$ and hence $H^{2}\sim2GM/a^{3}\to\infty$, an actual
big-bang singularity. The de~Sitter core replaces that constant $M$ by a mass
that scales away as $a^{3}$, and the divergence disappears at its root. The
enclosed mass is therefore \emph{smallest} at early times, and the
ADM mass $M$, fixed and finite throughout, never appears in the plateau
value \eqref{eq:inflate}. It fixes when the universe leaves the core through the transition scale $a_{\ast}^{3}\sim2GM\ell^{2}$, i.e.\ the number of
e-folds of the de~Sitter phase before the matter-like remainder takes over.

The two readings thus coincide on \cref{eq:inflate}: the vanishing of the
matter-like enclosed mass (energetic) is the same fact as the survival of the
core vacuum as the sole early source (dynamical). The resulting early phase is
non-singular in curvature and coasts to a constant $H$: de~Sitter coasting,
$a\propto e^{\Hinf t}$ with $q=-1$, an inflationary plateau instead of a linear
($q=0$) coasting law. We defer the accompanying statement about geodesic
completeness, which is the one respect in which this ``non-singularity'' must be
qualified, to \cref{ssec:coresurvival}.
\\
\\
Substituting the barotropic law $\rho\propto a^{-3(1+w)}$ into the McCrea--Milne
equation organises all regular-black-hole cosmologies by a single criterion. For $w\neq0$ the substitution $M\to\tfrac{4\pi}{3}\rho a^{3}$ is no longer backed by the
junction argument, which holds for dust, and its justification is the covariant one of
\cref{ssec:routeNderived}, where the same equation follows for any perfect fluid. The
modified Friedmann equation can be written $H^{2}=\tfrac{8\pi G}{3}\rho\,\mathcal{C}(a)$,
the regularising factor $\mathcal{C}(a)$ being read off by dividing the route-N
expansion rate by the bare barotropic one, e.g.  for Hayward with dust
$\mathcal{C}(a)=a^{3}/(a^{3}+2GM\ell^{2})$. Since the bare $\rho\propto a^{-3(1+w)}$
diverges at the origin, the induced
cosmology has  a finite, nonzero de~Sitter limit if and only if
\begin{equation}
  \mathcal{C}(a)\sim a^{3(1+w)}\qquad(a\to0),
  \label{eq:critexp}
\end{equation}
while a faster vanishing, $\mathcal{C}(a)\sim a^{p}$ with $p>3(1+w)$, is also
curvature-regular: $H^{2}\to0$ and all invariants vanish at the origin, a Minkowski
core of the kind carried by exponential mass functions $m\sim Me^{-\ell/r}$ and
matched to FLRW in~\cite{LiWuGe2025}. Only a slower vanishing, $p<3(1+w)$, is
singular. \Cref{eq:critexp} is thus the condition for a de~Sitter core, the case of
interest here, and not for regularity as such.
For dust ($w=0$) this requires the exponent $3$, which is nothing but the
de~Sitter-core condition $m\sim a^{3}$, whereas for radiation ($w=1/3$) it requires
exponent $4$. An important consequence of this criterion is the following: \emph{a metric can be a
perfectly regular black hole and still fail to regularise the cosmology}, and
whether it succeeds depends on the equation of state. Among our set, Bardeen, Jusufi--Singleton and NSS have precisely the exponent $3$ (dust-critical, with a de~Sitter core,
their regularisation scales, $g$, $\ell$ and $\sqrt{\theta}$ respectively, being held fixed 
and therefore independent of the source mass), while Hayward, Dymnikova and
Bonanno--Reuter are instead self-critical, reaching $3(1+w)$ for every $w>-1$ (for $w>-\tfrac13$ in the case of Bonanno--Reuter, whose linear term must stay below the mass term at the origin), because their cutoff terms ($2GM\ell^{2}$, $r_{0}^{3}=2GM\ell^{2}$ and
$\tilde\omega\gamma G_{0}^{2}M$ respectively) carry the source mass and therefore
inherit the density under the McCrea--Milne substitution.  For Jusufi--Singleton the core term of \cref{eq:mJS} is Bardeen's and the
self-energy term only enters at order $a^{5}$, so all six metrics regularise the dust
cosmology. The split between the two classes is a matter of convention: the Bardeen
profile becomes self-critical as soon as $g^{3}$ is tied to the mass, and Dymnikova is
listed as self-critical because her scale is defined with the mass in it,
$r_{0}^{3}=2GM\ell^{2}$ (\cref{tab:routeNcore}), which is the convention of our pipeline
and, as we will see in \cref{sec:covariant}, the one in which these metrics descend from
an action.

\section{Route N: solutions}
\label{sec:routeNsol}

Because route N gives $H^{2}$ as an explicit function of $a$, the cosmic time
follows from a single quadrature, namely $t(a)=\int \dd a/(a\sqrt{H^{2}})$. The status
of this integral is metric-dependent, and we now go through it. Here
 nothing is numerical from end to end: every model yields closed-form core
values and asymptotics, and only the time integral of the two transcendental
profiles requires numerics.

\subsection{Hayward}
With $A\equiv8\pi G\rho_{0}/3$ and $B\equiv A\ell^{2}$, $H^{2}=A/(a^{3}+B)$ and
\begin{equation}
  t(a)=\frac{2}{3\sqrt{A}}\Big[\sqrt{a^{3}+B}
  -\sqrt{B}\,\operatorname{arcsinh}\!\frac{\sqrt{B}}{a^{3/2}}\Big],
  \label{eq:taHaywardDust}
\end{equation}
exact for all $a$. For the radiation case ($\rho\propto a^{-4}$) one replaces $a^{3}\to a^{4}$ and the prefactor $\tfrac23\to\tfrac12$,
\begin{equation}
  t(a)=\frac{1}{2\sqrt{A}}\Big[\sqrt{a^{4}+B}
  -\sqrt{B}\,\operatorname{arcsinh}\!\frac{\sqrt{B}}{a^{2}}\Big].
  \label{eq:taHaywardRad}
\end{equation}
The limits are read directly: as $a\to0$ the inverse-hyperbolic term dominates
and inverts to $a\propto e^{t/\ell}$ (de~Sitter, $\Hinf=1/\ell$), whereas as $a\to\infty$ it drops out and $a\propto t^{2/3}$ (dust) or $t^{1/2}$ (radiation). Thus the
wide range of eras i.e. exponential core, smooth transition, power-law tail, is one
closed-form expression controlled by the single length $\ell$.

\subsection{Bardeen}
$H^{2}=2GM/(a^{2}+g^{2})^{3/2}$ gives a Gauss hypergeometric time,
\begin{equation}
  t(a)=\int\frac{(a^{2}+g^{2})^{3/4}}{a\sqrt{2GM}}\,\dd a
  =-\frac{a^{3/2}\Gamma(-\tfrac34)}{2\sqrt{2GM}\,\Gamma(\tfrac14)}\,
  {}_2F_1\!\Big(-\tfrac34,-\tfrac34;\tfrac14;-\tfrac{g^{2}}{a^{2}}\Big)+\text{const},
\end{equation}
closed-form but not elementary.

\subsection{Bonanno--Reuter}
$H^{2}=2G_{0}M/(a^{3}+\tilde\omega G_{0}a+\tilde\omega\gamma G_{0}^{2}M)$ is
rational, so $t(a)$ is elementary by partial fractions, but in terms of the three
(Cardano) roots of the cubic denominator.

\subsection{Dymnikova and NSS }
The exponential source $H^{2}=2GM(1-e^{-a^{3}/r_{0}^{3}})/a^{3}$ (Dymnikova) and
the incomplete-gamma source (NSS) admit no elementary or standard
special-function quadrature, so $t(a)$ is numerical. The analytic content is the
exact pair of asymptotics, de~Sitter core ($\Hinf=1/\ell$ for Dymnikova,
$\Hinf=[GM/(3\sqrt\pi\,\theta^{3/2})]^{1/2}$ for NSS) and dust tail, with a
numerically interpolated transition.

\subsection{Jusufi--Singleton}

With the mass function \eqref{eq:mJS}, route N gives
\begin{equation}
H^{2}(a)=\frac{2G\,m(a)}{a^{3}}
\;\xrightarrow{a\to0}\;\frac{2GM}{\ell^{3}}=\frac{\Lcore}{3},
\qquad
H^{2}(a)\;\xrightarrow{a\to\infty}\;
\frac{2G\mathcal{M}}{a^{3}}-\frac{G^{2}M^{2}}{a^{4}}+\mathcal{O}(a^{-5}),
\label{eq:JSroutenN}
\end{equation}
a de~Sitter core with the Bardeen value, the self-energy appearing only at order $a^{2}$
in the expansion about the core. At large $a$ the dust tail carries the ADM mass
$\mathcal{M}$ rather than the bare $M$, and the first correction is a negative $a^{-4}$
term with amplitude $M^{2}$. This is precisely the structure of the two-scale profile of
\cref{ssec:rung} below with $Q^{2}\to G^{2}M^{2}$, i.e.\ Jusufi--Singleton is a neutral
realisation of the decoupled charge--length deformation in which the ``charge''
amplitude is set by the mass itself. The time integral is numerical because of the
arctangent, as for the Gaussian cores. The self-energy bump of $\rhoMS$ makes this the
one metric of the set whose mean density can cease to decrease, which happens only for
$M/\ell>15$ (\cref{sec:criteria}), far from the core branch $M=\ell/2$ used in the
tables.

\begin{table}[t]
\centering
\renewcommand{\arraystretch}{1.9}
\begin{tabular}{|l>{$}l<{$}>{$}c<{$}c|}
\toprule
Metric & \text{Route N } H^{2}(a) & H^{2}(a\to0)=\Lcore/3 & $t(a)$ status \\
\midrule
Hayward        & \dfrac{2GM}{a^{3}+2GM\ell^{2}}            & 1/\ell^{2}        & elementary \\
Bardeen        & \dfrac{2GM}{(a^{2}+g^{2})^{3/2}}          & 2GM/g^{3}         & ${}_2F_1$ \\
Dymnikova      & \dfrac{2GM(1-e^{-a^{3}/r_{0}^{3}})}{a^{3}}& 2GM/r_{0}^{3}=1/\ell^{2} & numerical \\
NSS            & \dfrac{2GM}{a^{3}}\dfrac{\gamma(\frac32,\frac{a^{2}}{4\theta})}{\Gamma(\frac32)} & \dfrac{GM}{3\sqrt\pi\,\theta^{3/2}} & numerical \\
Bonanno--Reuter& \dfrac{2G_{0}M}{a^{3}+\tilde\omega G_{0}a+\tilde\omega\gamma G_{0}^{2}M} & \dfrac{2}{G_{0}\gamma\tilde\omega} & elem.\ up to roots \\
Jusufi--Singleton & \ \dfrac{2G\,m(a)}{a^{3}}\ \text{(\cref{eq:mJS})} & \ 2GM/\ell^{3} & \ numerical \\
\bottomrule
\end{tabular}
\caption{Route~N expansion rate, core value, and analytic status of the time
integral.  All six have a finite core. The arctangent in the Jusufi--Singleton mass function
puts its time integral with the Gaussian cores.}
\label{tab:routeNcore}
\end{table}

\subsection{Charged cores}
\label{ssec:rung}

The critical-exponent classification of \cref{eq:critexp} assigns each metric a
discrete series of barotropic indices, but in the neutral models these exponents are
locked: a single regularisation length controls the whole series, so no
individual term dominates a parametrically wide range of $a$. A charge supplies
a second, independent scale and breaks the degeneracy. That a charge can be
cosmologically active is not a new idea: a uniform net charge sources a stiff
($w=1$) background in the Proca framework~\cite{Barnes1979}, and a localised charge enters as an $a^{-4}$ ($w=1/3$) spectator in recent anisotropic $F(R)$
cosmology~\cite{Khan2026}. The present construction places both equations of
state on the same tower. We take the Ayón-Beato-García electric
metric~\cite{AyonBeato1998} (read through the charge-length duality, not
 as a literal net charge) and decouple its charge amplitude $Q$ from
its regularisation length $\ell$:
\begin{equation}
  f(r)=1-\frac{2GM\,r^{2}}{(r^{2}+\ell^{2})^{3/2}}
        +\frac{Q^{2}r^{2}}{(r^{2}+\ell^{2})^{2}},
  \qquad
  H^{2}_{N}(a)=\frac{2GM}{(a^{2}+\ell^{2})^{3/2}}-\frac{Q^{2}}{(a^{2}+\ell^{2})^{2}},
  \label{eq:ABGdecoupled}
\end{equation}
the single-scale ABG metric being the locked case $Q=\ell$. We stress that once
$Q$ and $\ell$ are decoupled the metric is no longer the Ayón-Beato--García
solution of nonlinear electrodynamics, and we do not know a field theory that sources it. Hence, \cref{eq:ABGdecoupled} is to be read as a phenomenological two-scale
profile whose only purpose here is to separate the terms of the series. The core value
follows from \cref{eq:corelimit} as $\Lcore/3=(2GM\ell-Q^{2})/\ell^{4}$, a
genuine de~Sitter core for $Q^{2}<2GM\ell$, a condition the component-isolation
limit below ($Q\to0$) satisfies with room to spare. The complementary regime
$Q^{2}>2GM\ell$ inverts the core to anti-de~Sitter: $H^{2}_{N}$ vanishes at a
minimal radius $a_{b}=[\,Q^{4}/(4G^{2}M^{2})-\ell^{2}\,]^{1/2}$, and the
cosmology bounces in place of emerging from an eternal de~Sitter past. This is
the over-charged, horizonless branch and lies outside the de~Sitter-core focus
of this work. We note only that a single map yields core survival and a
non-singular bounce on opposite sides of $Q^{2}=2GM\ell$, the bounce branch
being the more delicate (a radiation background overruns it unless the charge is
concentrated, $a_{b}\gg\ell$).

The large-$a$ expansion of \cref{eq:ABGdecoupled} exposes the series term by
term,
\begin{equation}
  H^{2}_{N}(a)=\underbrace{\frac{2GM}{a^{3}}}_{w=0}
  \;\underbrace{-\,\frac{Q^{2}}{a^{4}}}_{w=1/3}
  \;\underbrace{-\,\frac{3GM\ell^{2}}{a^{5}}}_{w=2/3}
  \;\underbrace{+\,\frac{2Q^{2}\ell^{2}}{a^{6}}}_{w=1}+\cdots,
  \label{eq:ladder}
\end{equation}
each power $a^{-n}$ corresponding through \cref{eq:routeNgeneric} to a fluid of
index $w=(n-3)/3$. We note that the $w=1/3$ term is carried by the charge $Q$ alone and the $w=2/3$ term by the regularisation length $\ell$ (and the mass) alone. The two cross at
\begin{equation}
  a_{c}=\frac{3GM\ell^{2}}{Q^{2}},
  \label{eq:across}
\end{equation}
below which the regularisation ($w=2/3$) term dominates the deviation from the
dust tail and above which the charge ($w=1/3$) term does. The window over which
the $w=2/3$ term leads, $\ell\lesssim a\lesssim a_{c}$, spans
$\ln(3GM\ell/Q^{2})$ e-folds and opens without bound as $Q\to0$, i.e. the limit in
which \cref{eq:ABGdecoupled} reduces to Bardeen and the deviation becomes purely
$w=2/3$. The electric and magnetic sectors thus sit on adjacent terms, and
decoupling the charge is exactly what exposes the magnetic one.

\begin{figure}[t]
  \centering
  \includegraphics[width=\textwidth]{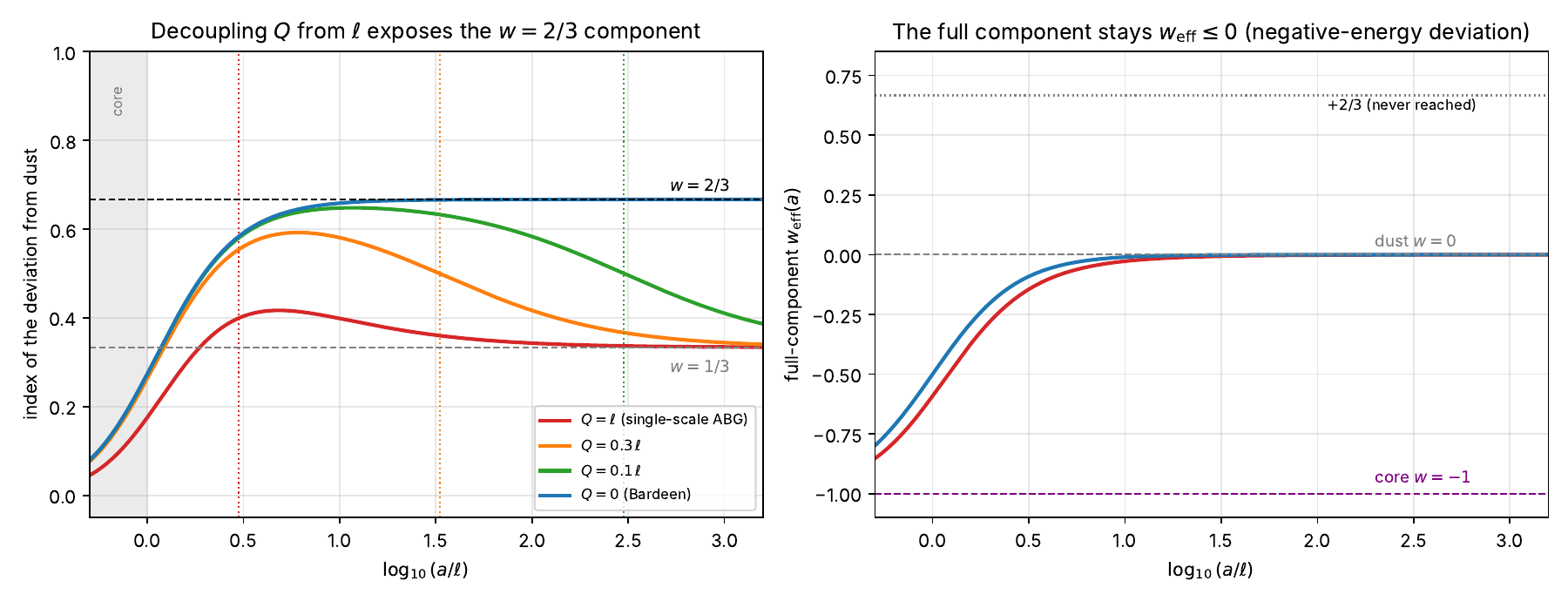}
  \caption{Isolating a single exponent with the two-scale metric
    \eqref{eq:ABGdecoupled} ($M=\ell=1$, $G=1$). \textbf{Left:} the barotropic
    index of the deviation from the dust tail as the charge $Q$ is decoupled
    from $\ell$. As $Q\to0$ the $w=2/3$ ($a^{-5}$) component dominates an ever-wider
    window $\ell\lesssim a\lesssim a_{c}=3GM\ell^{2}/Q^{2}$ (dotted lines: the
    crossover $a_{c}$), the $Q=0$ limit is Bardeen, whose deviation is purely
    $w=2/3$. \textbf{Right:} the \emph{full} effective equation of state
    $\weff(a)$ never reaches $+2/3$. Every route-N correction is a
    negative-energy enclosed-mass deficit, so $\weff$ runs from the dust value
    $0$ to the core value $-1$ and stays $\le0$. The $w=2/3$ component is the index of
    the deviation, not a positive fluid. }
  \label{fig:rung}
\end{figure}

Measured against the dust tail $H^{2}_{\rm dust}=2GM/a^{3}$, the fractional
deviation in the window is fixed at leading order by the regularisation length,
with an exact coefficient,
\begin{equation}
  \frac{\delta H^{2}}{H^{2}_{\rm dust}}\simeq-\frac{3\ell^{2}}{2a^{2}}
  \propto-(1+z)^{2},
  \qquad
  \delta H^{2}\propto a^{-5}\propto(1+z)^{5}.
  \label{eq:deviation}
\end{equation}
The two scalings carry different content. The \emph{absolute} deviation grows as
$(1+z)^{5}$, steeper than radiation and not degenerate with spatial curvature,
whose contribution to $H^{2}$ is $(1+z)^{2}$, so in a universe in which the route N fluid
is subdominant, this component would be identifiable in an
$H(z)$ fit instead of absorbable into $\Omega_{k}$. The sign plays a major role:
the full deviation from the dust tail is negative, $H_{N}^{2}(a)<2GM/a^{3}$ for all $a$,
because in route N the source is the
\emph{enclosed} Misner--Sharp mass, which both the regularisation and the
charge can only deplete (the positive $a^{-6}$ term in \cref{eq:ladder} is a
subleading part of this negative remainder, not a positive energy component).
This is the same monotonicity that underlies the crossing criterion
(\cref{sec:criteria}). The deviation is therefore a negative-energy
contribution, and the full effective equation of state of the metric component,
\begin{equation}
  \weff(a)=-1-\frac{1}{3}\frac{\dd\ln H^{2}_{N}}{\dd\ln a},
  \label{eq:weffrung}
\end{equation}
never reaches $+2/3$: for $Q\le\ell$ at the parameters of \cref{fig:rung} (in general for $Q^{2}\le\tfrac32GM\ell$), which covers the locked case and the whole
component-isolation limit, it runs monotonically from the dust value $\weff=0$ at
large $a$ to the core value $\weff=-1$, staying $\le0$ throughout
(\cref{fig:rung}). (For $\tfrac32GM\ell<Q^{2}<2GM\ell$ the Misner--Sharp
density of \cref{eq:ABGdecoupled} develops a shell and $\weff$ briefly undershoots
$-1$, exactly as the criterion of \cref{sec:criteria} demands. We do not use that regime.) What scales as $w=2/3$ is the index of the (negative)
deviation, equivalently the leading bend of $H(z)$ away from
$\Lambda$CDM-dust, not a standalone $w=2/3$ fluid epoch, which the same sign
that protects core survival forbids in route N. The classification is in this
sense observationally interpretable, each component being the leading $H(z)$ deviation in
a controllable redshift window. Route N itself, without baryons and radiation, is not a viable
late-time background, so the classification is structural and should not be read as a forecast.

\section{Route N: core survival and cosmography}
\label{sec:routeNcosmo}

\subsection{The de Sitter core survives}
\label{ssec:coresurvival}

In route N the question of whether the de~Sitter core ``survives'' (whether it
yields a non-singular, bounded-$H$ early phase) is well posed, because the regular
black hole is the sole source and there is no separate matter or radiation to diverge as $a\to0$. (In the covariant reading of \cref{ssec:routeNderived} the same expansion is sourced by dust, whose density does grow toward the past while the curvature stays bounded.) By \cref{eq:routeNgeneric} survival is controlled entirely by
the small-$a$ behaviour of $m(a)$: a de~Sitter core, $m\sim a^{3}$, gives
$H^{2}\to\Lcore/3$, a finite constant. We verify in addition that $H^{2}(a)$ is
\emph{bounded}: for all six de~Sitter-core metrics $\dd H^{2}/\dd a<0$ (for Jusufi--Singleton on its core branch, see \cref{sec:criteria}), so
$H^{2}\le H^{2}(0)=\Hinf^{2}$, the mean of a centre-peaked decreasing density
being itself decreasing. A bounded $H$ tending to a nonzero constant means there
is no curvature singularity and no big bang in cosmic time, which is the limiting-curvature behaviour of~\cite{FMM1989,MB1992,BMS1993} obtained here from the mass function rather than from a thin-shell or a Lagrange-multiplier construction: integrating
$\dd\ln a/\dd t=H$ gives $t\to-\infty$ as $a\to0$, an eternal de~Sitter past,
with $a(t)\to a_{i}e^{\Hinf t}$.

This ``non-singularity'' should be stated with care, because there are two
distinct notions and route N satisfies one but not the other. The model is \emph{curvature-regular}, i.e.\  all invariants are finite, $H$ is bounded, and there
is no big-bang singularity reached at finite cosmic time. It is not, however,
geodesically complete: the de~Sitter past in spatially flat slicing has
$a(t)\to a_{i}e^{\Hinf t}$, and the affine length of a past-directed radial null
geodesic, $\int^{t_{0}}\!a(t)\,\dd t\sim\int^{t_{0}}e^{\Hinf t}\dd t$, is finite,
so the model is past null incomplete. This is not a pathology peculiar to route
N. It is the generic feature of any de~Sitter or inflationary epoch, the same
past incompleteness that the Borde--Guth--Vilenkin analysis~\cite{BGV2003}
attaches to standard inflation, and the same conclusion reached for the minimal matched daughter by
Easson~\cite{Easson2026}, whose flat and open branches are likewise
``nonsingular in cosmic time and monotonic'' but past null incomplete. This does not undermine route N. It rather clarifies its interpretation, namely that the de~Sitter core is an \emph{inflationary} phase with the same geodesic structure inflation is known to have, and not a complete eternal past. The accompanying obstruction
theorem~\cite{Easson2026} states that a minimal, no-shell, asymptotically flat
\emph{matched} daughter cannot be at once indefinitely expanding,
curvature-regular, geodesically complete and consistent with the averaged null
energy condition. Route N realises the curvature-regular, expanding, inflationary member of this family. With the pressure sector fixed by the continuity equation
(\cref{sec:routeN}), its status inside the obstruction can be checked condition by condition.
 Writing $\mu=d\ln m/d\ln a$ (the deviation index of
\cref{ssec:cosmography}), the local and volume-averaged Misner--Sharp densities stand in
the ratio $\rho_{\mathrm{MS}}/\bar\rho_{\mathrm{MS}}=\mu/3$, so the effective
equation of state obeys
\begin{equation}
1+w_{\mathrm{eff}}\;=\;1-\frac{\rho_{\mathrm{MS}}}{\bar\rho_{\mathrm{MS}}}\,.
\label{eq:necidentity}
\end{equation}
The null energy condition of the matched FLRW region is thus the statement
that the local density never exceeds its own volume average. For a
monotonically decreasing profile this is automatic: the mean of a
decreasing function always exceeds its endpoint value, with saturation only
in the $r\to0$ limit, where the core is exactly de Sitter and the condition is
marginally met, as a de Sitter phase should meet it. Indefinite expansion,
curvature regularity and the averaged null energy condition therefore hold
verifiably for every monotone core, and the obstruction does not merely permit
but forces the failure of the one remaining condition, geodesic completeness. The
incompleteness is the familiar one of Borde, Guth and Vilenkin, shared by any inflationary phase. This is in full agreement with~\cite{Easson2026}, and the condition that fails has been identified and not assumed.

\begin{figure}[t]
  \centering
  \includegraphics[width=\textwidth]{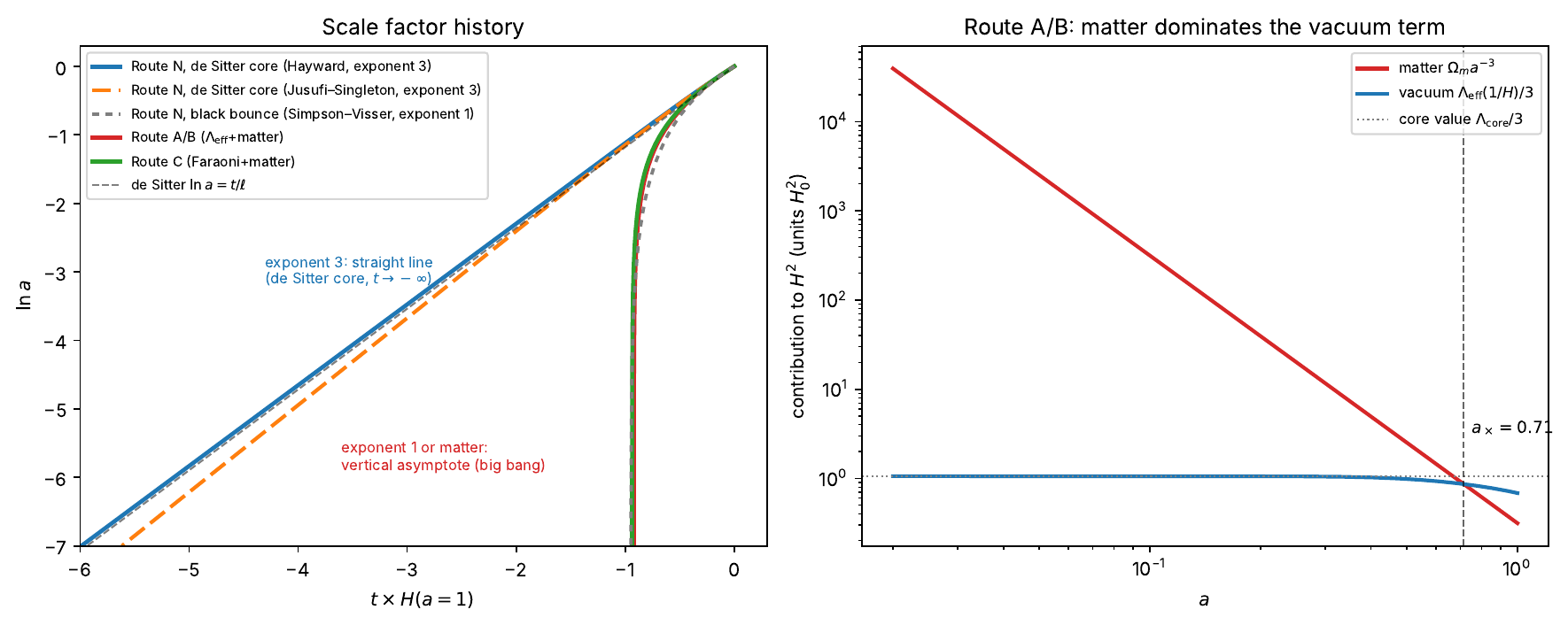}
  \caption{ \textbf{Left:} scale-factor history $(t,\ln a)$ in route N for two
    de~Sitter cores (Hayward, solid, and Jusufi--Singleton, long dashes, both exponent $3$: straight lines, $t\to-\infty$, no
    singularity) and for the Simpson--Visser lapse function $f=1-2GM/\sqrt{r^{2}+\ell^{2}}$ read in the gauge of \cref{eq:ssmetric}, with $r$ as the areal radius (short dashes,
    exponent $1$: a vertical asymptote, a big bang at finite time), versus the
    Hubble-radius routes with ordinary matter present (also a vertical asymptote).
    The singularity in route N is a property of the mass-function exponent, not of
    the route. \textbf{Right:} in the Hubble-radius routes the regular-core vacuum saturates to the finite value $\Lcore/3$, which the diverging matter exceeds before $a_{\times}$, which is why core survival is a feature specifically
    of route~N, where the regular black hole is the entire content.}
  \label{fig:coresurvival}
\end{figure}

These observations combine into a simple
statement. If $f=1-2Gm(r)/r$ defines a regular black hole with a de~Sitter core,
$m(r)\sim r^{3}/(2G\ell_{\star}^{2})$ as $r\to0$, then the route-N cosmology
\eqref{eq:routeNgeneric} is curvature-regular with no big-bang singularity: $H^{2}$
is bounded, $H^{2}(a\to0)=1/\ell_{\star}^{2}$, and $a(t)\to a_{i}e^{t/\ell_{\star}}$,
while the de~Sitter past is inflationary and, like any such phase, past null
incomplete.  This holds for all six metrics of \cref{tab:routeNcore}. It would fail for a mass function of exponent one, $m\sim a$ near the origin, and
\cref{fig:coresurvival} shows the case of the Simpson--Visser lapse function~\cite{SimpsonVisser2019} read in the gauge of \cref{eq:ssmetric} for contrast (the black-bounce geometry itself has areal radius $\sqrt{r^{2}+\ell^{2}}$ and is not of that form) and, for comparison, the Hubble-radius routes (in which the core appears instead as a bounded vacuum term that ordinary matter dominates toward the past).

\subsection{A validation case: the quantum Oppenheimer--Snyder metric}
\label{ssec:qos}

Route N can be tested against a case in which the answer is known beforehand.
Lewandowski, Ma, Yang and Zhang~\cite{LMYZ2023} applied the effective dynamics of
loop quantum cosmology~\cite{APS2006,APS2006b} to Oppenheimer--Snyder collapse and obtained the exterior
\begin{equation}
  f(r)=1-\frac{2GM}{r}+\frac{\alpha G^{2}M^{2}}{r^{4}},
  \qquad \alpha=16\sqrt{3}\,\pi\gamma^{3}\ell_{\rm Pl}^{2},
  \label{eq:qos}
\end{equation}
with $\gamma$ the Barbero--Immirzi parameter. This geometry is not a regular
black hole. Its Kretschmann scalar diverges as $r^{-12}$ and $f\to+\infty$ as
$r\to0$, so there is no de~Sitter core. It is nevertheless the ideal test of
\cref{eq:routeNgeneric}, because it is built by matching a spatially flat FLRW
dust ball to a static exterior across a timelike surface. The daughter Friedmann
equation is therefore known independently, from the quantum theory rather than
from our construction.

Reading the mass function off \cref{eq:qos} gives $m(r)=M-\alpha GM^{2}/2r^{3}$.
Route N then yields
\begin{equation}
  H^{2}=\frac{2Gm(a)}{a^{3}}
       =\frac{8\pi G}{3}\rho-\frac{16\pi^{2}G^{2}\alpha}{9}\rho^{2}
       =\frac{8\pi G\rho}{3}\Big(1-\frac{\rho}{\rho_{c}}\Big),
  \qquad \rho_{c}=\frac{3}{2\pi G\alpha},
  \label{eq:qoslqc}
\end{equation}
where we used $M=\tfrac{4\pi}{3}\rho a^{3}$. This is the effective Friedmann
equation of loop quantum cosmology. The agreement is not only in form. Inserting
the value of $\alpha$ from \cref{eq:qos} and the area gap
$\Delta=4\sqrt{3}\pi\gamma\ell_{\rm Pl}^{2}$ into the loop quantum cosmology
expression $\rho_{c}=3/8\pi G\gamma^{2}\Delta$ gives
\begin{equation}
  \rho_{c}=\frac{\sqrt{3}}{32\pi^{2}G\gamma^{3}\ell_{\rm Pl}^{2}},
  \label{eq:rhoc}
\end{equation}
which is exactly the value returned by \cref{eq:qoslqc}, including the numerical coefficient.
Route N therefore reproduces a result derived from loop quantum gravity, in the
one case in this paper where an external check is available.

The case is instructive for a second reason. Because $m$ crosses zero at
$a_{b}=(\alpha GM/2)^{1/3}$, the route N cosmology of \cref{eq:qos} does not
approach a plateau. It bounces. This identifies two distinct ways in which route
N can avoid a big bang, and only one of them is realised by the metrics of this
paper,
\begin{itemize}\setlength\itemsep{2pt}
  \item \emph{regularisation by plateau}, when $m\to0^{+}$ as $a^{3}$, giving the
        finite $H^{2}\to\Lcore/3$ of \cref{eq:inflate}, and
  \item \emph{regularisation by bounce}, when $m$ changes sign, giving $H^{2}=0$
        at finite $a_{b}$ and no de~Sitter phase at all.
\end{itemize}
The six metrics of \cref{tab:routeNcore} belong to the first
class. The quantum
Oppenheimer--Snyder metric belongs to the second, and indeed it is not a
de~Sitter-core metric.

Finally, this metric separates the three routes in a way the six regular models
do not. Its local Misner--Sharp density is
$\rhoMS=3G\alpha M^{2}/8\pi r^{6}$, so
$\Leff(r)=3G^{2}\alpha M^{2}/r^{6}$ diverges as $r\to0$ rather than saturating.
Route A/B, which reads $\Leff$, therefore sees no core whatsoever, while route N,
which reads the mean density $\bar\rho_{\rm MS}$, sees a bounce. The two
prescriptions give qualitatively different early universes from a single metric.
Route C is unaffected, since its source $2GH^{3}m(1/H)=2GMH^{3}-G^{2}\alpha
M^{2}H^{6}$ still decays as $H^{3}$, the enclosed mass saturating at $M$ as it
does for every metric in \cref{tab:routeC}. We note in passing that $\rhoMS$ here
is monotone, so the crossing criterion of \cref{sec:criteria} applies unchanged
and this metric cannot cross the phantom divide either. Since
$\alpha\sim\ell_{\rm Pl}^{2}$, none of this bears on the late-time data.
 
\subsection{Cosmography: one function controls everything}
\label{ssec:cosmography}

The kinematics of route N collapse onto a single running quantity, the
logarithmic slope of the mass function, taken as
\begin{equation}
  \mu(a)\equiv\frac{\dd\ln m}{\dd\ln a}=\frac{a\,m'(a)}{m(a)},
\end{equation}
which decreases monotonically from $3$ (de~Sitter core, $m\sim a^{3}$) to $0$
(dust, $m\to M$). The kinematic parameters follow from this single function. The
standard cosmographic definitions are
$q=-\ddot a a/\dot a^{2}=-1-\dot H/H^{2}$ and $j=\dddot a/(aH^{3})$, the deceleration and jerk~\cite{Visser2004}. Converting time derivatives to
derivatives in $N\equiv\ln a$ via $\dd/\dd t=H\,\dd/\dd N$ gives the useful
identities $\dot H=\tfrac12\,\dd H^{2}/\dd N$ and
$\ddot H=\tfrac12 H\,\dd^{2}H^{2}/\dd N^{2}$. Since $H^{2}=2Gm/a^{3}$ depends on
$a$ only through $m$, we have $\dd\ln H^{2}/\dd N=\mu-3$ and, differentiating
once more, $\dd^{2}\ln H^{2}/\dd N^{2}=\mu'$ and then substituting into the definitions
and using $\weff$ from the continuity equation
$\dot\rho_{\rm eff}=-3H(1+\weff)\rho_{\rm eff}$ with $\rho_{\rm eff}=3H^{2}/8\pi G$,
the deceleration parameter, effective equation of state and jerk reduce to
\begin{equation}
  q(a)=\frac{1-\mu}{2},\qquad
  \weff(a)=-\frac{\mu}{3},\qquad
  j(a)=1+\tfrac12\big(\mu^{2}-3\mu+\mu'\big),
  \label{eq:cosmography}
\end{equation}
with $\mu'\equiv\dd\mu/\dd\ln a$, so $\mu$ describes the cosmographic parameters.
 The first two are algebraic and only the
jerk feels the running. At the core ($\mu\to3$) one has $q\to-1$, $\weff\to-1$,
$j\to1$, and in the dust limit ($\mu\to0$) one has $q\to+\tfrac12$, $\weff\to0$,
$j\to1$. Thus $j=1$ at both ends and any departure is a transient feature of the
transition: a discriminator, since $\Lambda$CDM has $j\equiv1$ at all
epochs.

\begin{figure}[t]
  \centering
  \includegraphics[width=\textwidth]{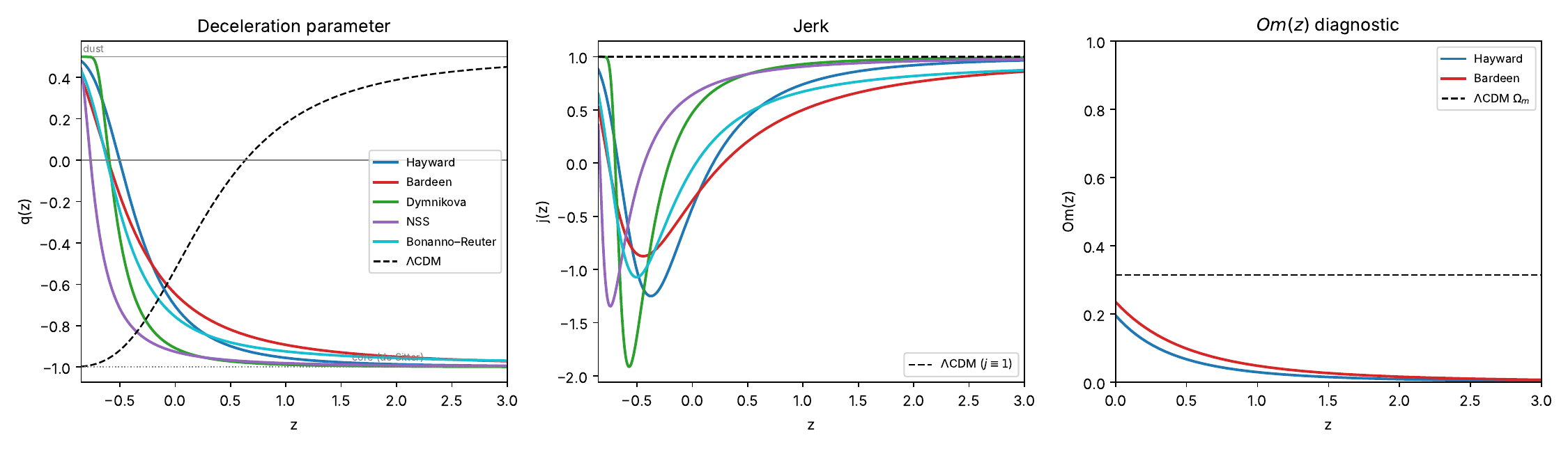}
  \caption{Route N cosmography. \textbf{Left:} the deceleration parameter runs
    from $q=-1$ (de~Sitter core, in the past) to $q=+\tfrac12$ (dust, in the
    future), the reverse sense to $\Lambda$CDM, which the curves cross.
    \textbf{Centre:} the jerk equals unity for $\Lambda$CDM at all $z$, while
    route N departs from it transiently and always from below (today the jerk is negative for the rational cores and positive for Dymnikova and NSS). \textbf{Right:} the $Om(z)$ diagnostic decreases
    with $z$ (Hayward: $Om=1/[1+2GM\ell^{2}(1+z)^{3}]$), against the constant
    $\Omega_{m}$ of $\Lambda$CDM. Jusufi--Singleton, nearly degenerate with Bardeen (\cref{tab:routeNindicators}), is not drawn.}
  \label{fig:cosmographyfig}
\end{figure}

The control function is closed-form for the rational metrics, e.g.\
$\mu_{\rm Hayward}=6GM\ell^{2}/(a^{3}+2GM\ell^{2})$ and
$\mu_{\rm Bardeen}=3g^{2}/(a^{2}+g^{2})$. A second model-independent diagnostic is
the $Om(z)$ function of Sahni, Shafieloo and Starobinsky~\cite{Sahni2008},
\begin{equation}
  Om(z)\equiv\frac{E^{2}(z)-1}{(1+z)^{3}-1},\qquad E(z)\equiv\frac{H(z)}{H_{0}},
  \label{eq:Omdef}
\end{equation}
constructed so that $Om(z)=\Omega_{m}$ is a constant for $\Lambda$CDM, with any
slope signalling a departure. In route N, $E^{2}(a)=m(a)/[a^{3}m(1)]$ from
\cref{eq:routeNgeneric}, and for Hayward \cref{eq:Omdef} collapses to the closed
form $Om(z)=1/[1+2GM\ell^{2}(1+z)^{3}]$, a decreasing function of $z$ , the
signature that the matter fraction was smaller in the past, because the past was
core-dominated. \Cref{tab:routeNindicators} lists the present-day indicators at
fiducial scales, and \cref{fig:cosmographyfig} shows $q(z)$, $j(z)$ and $Om(z)$
against $\Lambda$CDM.

\begin{table}[t]
\centering
\renewcommand{\arraystretch}{1.3}
\begin{tabular}{|l|c|c|c|c|c|}
\toprule
Metric & $\mu_{0}$ & $q_{0}$ & $w_{0}$ & $j_{0}$ & $z(q{=}0)$ \\
\midrule
Hayward          & 2.411 & $-0.706$ & $-0.804$ & $-0.420$ & $-0.50$ \\
Bardeen          & 2.292 & $-0.646$ & $-0.764$ & $-0.352$ & $-0.61$ \\
Jusufi--Singleton & 2.314 & $-0.657$ & $-0.771$ & $-0.320$ & $-0.62$ \\
Dymnikova        & 2.816 & $-0.908$ & $-0.939$ & $+0.472$ & $-0.60$ \\
NSS              & 2.852 & $-0.926$ & $-0.951$ & $+0.643$ & $-0.77$ \\
Bonanno--Reuter  & 2.513 & $-0.757$ & $-0.838$ & $-0.053$ & $-0.62$ \\
\midrule
$\Lambda$CDM     & --- & $-0.528$ & $-1$ & $+1.000$ & $+0.63$ \\
\bottomrule
\end{tabular}
\caption{Route N present-day cosmographic indicators at fiducial scales
($M=\ell/2$, where each metric's regularisation scale $\ell$ takes its
Table~\ref{tab:cpl} value), with $\Lambda$CDM ($\Omega_{m}=0.315$) for comparison. The sign
of $z(q{=}0)$ is negative for route N (acceleration ends in the future) and
positive for $\Lambda$CDM (it began in the past), $j_{0}\ne1$ is a clear discriminator. Jusufi--Singleton sits next to Bardeen, from which it differs only through the self-energy term.}
\label{tab:routeNindicators}
\end{table}

\subsection{Late-time behaviour and the meaning of the result}

For large $a$ the mass saturates, $m\to M$ and we get $H^{2}\to 2GM/a^{3}$, i.e. the Einstein--de Sitter universe with
$q\to \frac{1}{2}$, $w_{\rm eff} \to 0$ and $a \propto t^{2/3}$. The Hubble parameter
is then positive at all times and there is no recollapse. There is also no late-time dark energy behaviour. The single fluid we started with has simply become dust (as can be seen from the exponent). The
deceleration parameter therefore runs the wrong way round for a dark-energy model (the de Sitter phase
lies in the past, not the future), and it is more natural to read route N as a model of inflation with a graceful exit, in which the regularisation length
$\ell$ takes over the role usually played by an inflaton potential. As a model of the present acceleration
it does not work, and this is not a question of tuning. It would need the universe
to accelerate at high redshift, whereas nucleosynthesis, the microwave background and structure formation
all require a matter-dominated era there. The observables that distinguish it from
$\Lambda$CDM are the sign of $\dd q /\dd z$ and the jerk.

It is also worth noting that adding a bare cosmological constant does not rescue the late-time
behaviour, namely $H^{2}=2Gm(a)/a^{3}+\Lambda/3$. The reason is that the regular-black-hole
 term is bounded, $2Gm(a)/a^{3}\le1/\ell^{2}$,
so the two scales conflict: at the inflationary scale $\ell\sim10^{4}\,\ell_{\rm Pl}$ the core is left at
$a_{\ast}^{3}\sim 2GM\ell^{2}$, which for any astrophysical $M$ lies far before nucleosynthesis, so that
by today the term has redshifted as $a^{-3}$ from a Planckian value and is negligible against $\Omega_{m,0}H_{0}^{2}$, while forcing it to
carry $\Omega_{m,0}$ requires $\ell\sim H_{0}^{-1}$, at which scale it no longer
scales as $a^{-3}$ across the observed range (deviating from dust by tens of
percent already at $z\simeq0.5$) and the model fails the BAO $H(z)$ test. A bare
$\Lambda$ adds late acceleration but cannot repair a matter sector that is the
wrong shape. The physically consistent picture keeps the regular core as the
inflaton and supplies late-time matter and acceleration by the standard
$\Lambda$CDM ingredients: three independent scales, with no unification
claimed.

The inflationary reading is delimited by three qualifications. First, the e-fold count is set by
the exit but not the entry: the de Sitter past is eternal, so the number of e-folds from any
initial $a_{i}$ to the transition $a_{\ast}^{3}\sim 2GM\ell^{2}$ is $N\simeq \frac{1}{3}\ln(a_{\ast}^{3}/a_{i}^{3})$,
formally unbounded, and the ADM mass fixes only when the exit occurs. Second,
the exit is to dust and not to radiation, i.e. we claim no reheating mechanism and no thermal history
compatible with nucleosynthesis. Third, we compute no perturbation spectrum. An inflationary background of the same origin, a brane moving in a regular black-hole
bulk, is studied in \cite{Sueto2026}.

\section{The covariant origin of route N}
\label{sec:covariant}

Routes N, A/B and C were introduced as prescriptions. We now ask which of
them, if any, is the cosmological sector of a covariant theory whose static
vacuum solution is the regular black hole we started from. We will see that within the quasi-topological class the question can be answered, and the answer is route N.

\subsection{Quasi-topological gravities and their two sectors}
\label{ssec:qtg}

Quasi-topological gravities are higher-curvature theories built so that
their field equations remain second order on warped-product backgrounds. In
$D\geq5$ they can be polynomial in the curvature \cite{BuenoCano2025}, in
four dimensions they must be non-polynomial \cite{BorissovaCarballoRubio2026,BCHM2026,Borissova2026dilaton}. On the static line element \eqref{eq:ssmetric} the
vacuum equations of the whole class reduce to a single algebraic relation
for the function $\psi\equiv(1-f)/r^{2}$,
\begin{equation}
h(\psi)=\frac{2GM}{r^{3}},
\label{eq:hstatic}
\end{equation}
where the characteristic function $h$ is fixed by the couplings of the
theory, $h(\psi)=\psi+\sum_{n\geq2}\alpha_{n}\psi^{n}$ for a polynomial tower
and an arbitrary function~\cite{BorissovaMagueijo2026} with $h(\psi)\simeq\psi$ at small $\psi$ in the
non-polynomial case, and $M$ is an integration constant, the ADM mass. A
regular black hole with a de Sitter core corresponds to an $h$ that diverges
at the finite value $\psi_{0}=\Lcore/3$, so that $\psi\to\psi_{0}$ as $r\to0$.

The same theories have an FLRW sector, and Borissova and
Magueijo \cite{BorissovaMagueijo2026} have shown that for a perfect-fluid
source it reduces to one algebraic relation as well,
\begin{equation}
h\!\left(H^{2}+\frac{k}{a^{2}}\right)=\frac{8\pi G}{3}\,\rho ,
\label{eq:hcosmo}
\end{equation}
with the same $h$ as in Eq.~\eqref{eq:hstatic}. Two conditions are forced
along the way. The $tr$ equation imposes the integrability condition that
defines the quasi-topological class, and the perfect-fluid source sets to
zero the second generating function that non-polynomial theories can carry in
general, so that the cosmology is governed by $h$ alone. Equation
\eqref{eq:hcosmo} is the generalised Friedmann equation of the theory, with
the continuity equation unchanged.

\subsection{Which metrics admit a characteristic function}
\label{ssec:hexists}

Not every regular metric descends from a theory of this kind. The couplings
in $h$ are constants of the theory, so $h$ must not depend on the mass $M$,
which is an integration constant. Writing $X\equiv 2GM/r^{3}$, Eq.
\eqref{eq:hstatic} says that $\psi$ must be a function of $X$ alone. For a
metric given as $f(r;M,s)$ with regularisation scale $s$ this is the
condition that the Jacobian
\begin{equation}
\frac{\partial(\psi,X)}{\partial(r,M)}=0
\label{eq:jacobian}
\end{equation}
vanish identically, which can be checked without inverting $\psi(r)$ and
is therefore usable for the transcendental profiles too. We have applied it
to the six metrics of Section~\ref{sec:metrics}, and the outcome depends on
whether the regularisation scale is held fixed or carries the mass.

With the scales as written in Section~\ref{sec:metrics} and Table~\ref{tab:routeNcore},
Hayward and Dymnikova pass. For Hayward
\begin{equation}
h_{\rm Hayward}(\psi)=\frac{\psi}{1-\ell^{2}\psi}
=\psi+\ell^{2}\psi^{2}+\ell^{4}\psi^{3}+\dots,
\label{eq:hHayward}
\end{equation}
the geometric tower $\alpha_{n}=\ell^{2(n-1)}$ already noted in
\cite{BuenoCano2025}. For Dymnikova, whose scale $r_{0}^{3}=2GM\ell^{2}$
carries the mass by construction, $\psi=X\,[1-e^{-1/(\ell^{2}X)}]$ gives
$h=\psi$ up to corrections exponentially small in $1/\psi$, which is the
statement of Konoplya and Zhidenko \cite{KonoplyaZhidenko2024} that the
Dymnikova tower is non-perturbative in the coupling, and the explicit
four-dimensional theory of \cite{BorissovaCarballoRubio2026}, whose
characteristic function involves the Lambert function, inverts this very relation (the two agree numerically). Bardeen, NSS and
Jusufi--Singleton fail because $g$, $\theta$ and the bare scale enter $\psi$ in
a way that cannot be absorbed into $X$. Bardeen and NSS pass if their
regularisation volume is tied to the mass, $g^{3}=2GM\ell^{2}$ and
$\theta^{3/2}\propto GM$,
$\ell$ being then the coupling of the theory, and the resulting characteristic
functions are non-polynomial: for Bardeen, $\psi=X\,[1+(\ell^{2}X)^{2/3}]^{-3/2}$
gives $h=\psi+\tfrac32\ell^{4/3}\psi^{5/3}+\dots$, with fractional powers.
Jusufi--Singleton passes for no scaling of its bare scale. Its core term would need $\ell^{3}\propto M$, but the self-energy term of \cref{eq:mJS} then enters $\psi$ with amplitude $G^{2}M^{2}/\ell^{4}\propto M^{2/3}$ and still carries the mass, so that the Jacobian \eqref{eq:jacobian} is non-zero for every power law $\ell\propto M^{p}$.
Bonanno--Reuter passes for no scaling of its
parameters at all: the term linear in $r$ in the denominator of its $\psi$,
proportional to $\tilde\omega G_{0}$, cannot be written through $X$, and the
metric becomes quasi-topological only in the limit $\tilde\omega\to0$ at
fixed $\gamma\tilde\omega$, where it collapses to Hayward with
$2\ell^{2}=\gamma\tilde\omega G_{0}$. With the literature values of
Section~\ref{sec:metrics} the Bonanno--Reuter metric has no action in this class, which is consistent with its origin as an improvement
of the Schwarzschild solution rather than as a solution of anything. Its two terms belong to
different regimes. The linear one reproduces the large-distance running
$G(r)\simeq G_{0}(1-\tilde\omega G_{0}/r^{2})$, with $\tilde\omega$ fixed in \cite{BonannoReuter2000}
by matching to the one-loop result of the effective field theory of gravity \cite{Donoghue1994,HamberLiu1995},
and the mass-dependent one is the Hayward core. The effective field theory behind the first has
an action but is perturbative in the curvature and cannot produce a core, the tower behind the
second produces the core and not the tail, and the metric interpolates between the two.

The mass-dependent convention is not an artifice, it is the one in which Dymnikova wrote her metric, $r_{0}^{3}=r_{g}r_{\star}^{2}$ with $r_{g}=2GM$,
and it is the convention in which the quasi-topological constructions of
these metrics are written. Under it Bardeen and NSS would join Hayward and
Dymnikova in the self-critical class of Section~\ref{sec:routeN}, since their
cutoff terms would then carry the source mass. This is the first consequence
of the covariant reading: self-criticality, the property of regularising every
barotropic cosmology with $w>-1$, is the signature of a metric that
descends from a quasi-topological theory, and dust-criticality is the
signature of a scale that has been held fixed by hand.

\subsection{Route N is the FLRW sector}
\label{ssec:routeNderived}

Let us take a metric $f$ that passes the test, with characteristic function
$h$. Fill the universe with dust of density $\rho=3M/4\pi a^{3}$, the
comoving normalisation of Eq.~\eqref{eq:routeNgeneric}, and set $k=0$. Then
Eq.~\eqref{eq:hcosmo} reads $h(H^{2})=2GM/a^{3}$, which is Eq.
\eqref{eq:hstatic} with $r$ replaced by $a$. Its solution is therefore
$H^{2}=\psi(a)$, i.e.
\begin{equation}
H^{2}(a)=\frac{1-f(a)}{a^{2}}=\frac{2G\,m(a)}{a^{3}},
\label{eq:routeNderived}
\end{equation}
which is Eq.~\eqref{eq:routeNgeneric} exactly, for all $a$ and not only in
the two asymptotic regimes. Route N is the cosmological sector of the theory
whose vacuum is the regular black hole. The McCrea--Milne substitution of
Section~\ref{sec:routeN} and the Darmois matching of Eq.~\eqref{eq:junction}
are two further ways of saying the same thing, and the Oppenheimer--Snyder
collapse of \cite{BCHMV2025}, in which the interior dust ball obeys
\eqref{eq:hcosmo} and matches the regular exterior, is the dynamical version
of the statement. What the present derivation adds is that the equation is
not inherited from a junction but produced by the field equations on a
homogeneous background, so that the Kantowski--Sachs caveat of
Section~\ref{sec:routeN} no longer applies: the FLRW universe of Eq.
\eqref{eq:routeNderived} is a solution of the theory in its own right.

 With radiation instead of dust,
$\rho\propto a^{-4}$, the same $h$ gives for Hayward
$H^{2}=X/(1+\ell^{2}X)$ with $X=8\pi G\rho/3$, bounded by $1/\ell^{2}$, and
$H^{2}=X[1-e^{-1/(\ell^{2}X)}]$ for Dymnikova, bounded by the same value. The
self-critical regularisation of Section~\ref{sec:routeN} is thus a property
of the theory and not of the substitution, and the de Sitter emergence with
pure radiation found in \cite{BorissovaMagueijo2026} is this case. In this case the matter density itself grows without bound toward the past, $\rho\propto a^{-3}$ or $a^{-4}$, while the
curvature stays finite, a feature already noted there. In \cite{BorissovaMagueijo2026} this divergence is placed at infinite affine distance, which is the statement for observers comoving with the fluid, for whom it lies at $t\to-\infty$. Along null geodesics the same boundary is reached at finite affine parameter, which is the past incompleteness discussed in Section~\ref{ssec:coresurvival}. Moreover, the cosmography of Section~\ref{ssec:cosmography}, the critical-exponent
classification and the core-survival statement of
Section~\ref{ssec:coresurvival} are, for the metrics that pass the test,
statements about a covariant theory. The inflationary reading of route N
rests on a Lagrangian, even if the Lagrangian is non-polynomial and we make
no claim about its perturbative sector.

\subsection{On routes A/B and C}
\label{ssec:ABCnotderived}

The Hubble-radius routes can be written in the same algebraic form. Dropping
radiation, Eq.~\eqref{eq:routeABfriedmann} is
$\tilde h_{\rm AB}(H^{2})=8\pi G\rho_{m}/3$ with
\begin{equation}
\tilde h_{\rm AB}(\psi)=\psi-\frac{\Leff(\psi^{-1/2})}{3},
\qquad
\tilde h_{\rm C}(\psi)=\psi-2G\,\psi^{3/2}\,m(\psi^{-1/2}),
\label{eq:htilde}
\end{equation}
and Eq.~\eqref{eq:routeCfriedmann} is the analogous relation with
$\tilde h_{\rm C}$. Since the non-polynomial class admits an arbitrary
characteristic function, a theory with $\tilde h_{\rm AB}$ or
$\tilde h_{\rm C}$ exists. It is not, however, the theory of the black hole.
For Hayward one finds from Eqs.~\eqref{eq:routeABfriedmann} and
\eqref{eq:routeCfriedmann}
\begin{equation}
\frac{\partial\tilde h_{\rm AB}}{\partial M}
=-\frac{8M\ell^{2}\psi^{3}}{(1+2M\ell^{2}\psi^{3/2})^{3}},
\qquad
\frac{\partial\tilde h_{\rm C}}{\partial M}
=-\frac{2\psi^{3/2}}{(1+2M\ell^{2}\psi^{3/2})^{2}},
\label{eq:dhdM}
\end{equation}
neither of which vanishes. In routes A/B and C the black-hole mass has
become a coupling of the characteristic function, so that the static vacuum
of the corresponding theory carries a second, independent mass as its
integration constant and is not the regular black hole one began with. This
is the precise sense in which the two routes are phenomenological. They are
cosmologies of some quasi-topological theory, with a running law inherited
from a regular-black-hole family, but the link between that family and the
cosmology is broken at the level of the action. We regard this as a feature
of the comparison rather than a defect: routes A/B and C measure how far a
late-time running law can be pushed once the black hole is used as a
generating profile, and Section~\ref{sec:routeCimpl} quantifies the spread.

\subsection{A scalar is not the covariant completion}
\label{ssec:scalarlemma}

One might hope to obtain the same correction to $H^{2}$ from a running
Newton constant rather than from higher curvature. With dust the Hayward
equation \eqref{eq:routeNhayward} can indeed be written
$H^{2}=(8\pi G/3)\rho\,[1+8\pi G\ell^{2}\rho/3]^{-1}$, i.e.\ with a density
screened Newton constant, and on the static side the same metric is
$G_{\rm eff}(r)=G\,r^{3}/(r^{3}+2GM\ell^{2})$, which is the Bonanno--Reuter
form with $\tilde\omega\to0$. A dynamical scalar cannot do this. In a Brans--Dicke-type
scalar-tensor theory (coupling $\omega(\phi)$ and an arbitrary potential, which drops out of the combination below) in the Jordan frame, the combination of the $tt$ and
$rr$ field equations on the line element \eqref{eq:ssmetric}, where
$G^{t}{}_{t}=G^{r}{}_{r}$ identically, gives
\begin{equation}
f\left[\phi''+\frac{\omega(\phi)}{\phi}\,\phi'^{2}\right]
=-\,8\pi\,(\rho+p_{r}),
\label{eq:stlemma}
\end{equation}
so that with the radial equation of state $p_{r}=-\rho$ of
Section~\ref{ssec:Leffvalid}, or in vacuum, $(\phi^{\omega+1})''=0$ for
constant $\omega$ ($(\ln\phi)''=0$ if $\omega=-1$, with the same conclusion). The scalar is then linear in $r$ to the power
$1/(\omega+1)$ and cannot approach a finite, non-zero constant at infinity unless it is
constant everywhere.  The ansatz $g_{tt}g_{rr}=-1$, which every metric of
Section~\ref{sec:metrics} satisfies and on which the identification
$\Leff=8\pi G\rhoMS$ rests, is therefore incompatible with a non-trivial
Brans--Dicke scalar, and a scalar-tensor reconstruction of the route-N
expansion history would deliver the cosmology while losing the black hole.
The correction to $H^{2}$ in Eq.~\eqref{eq:hHayward} is a curvature
correction, $H^{2}+\ell^{2}H^{4}+\dots=8\pi G\rho/3$, and on the dust shell
the two readings coincide only because $\rho$ and $H^{2}$ are tied.

In summary, for Hayward and Dymnikova as written, and for Bardeen and NSS in the
convention in which their regularisation volume scales with the mass, route N
is the flat FLRW
sector of a four-dimensional pure-gravity theory, with the same couplings
that make the static solution regular. Its de Sitter core, its
self-critical regularisation and its cosmography are consequences of that
theory. Bonanno--Reuter and Jusufi--Singleton, the latter through its
self-energy term, have no origin in this class with their published
parameters, and their route-N cosmologies remain at the level of the
matching argument of Eq.~\eqref{eq:junction}. Routes A/B and C are phenomenological in a definite sense, i.e.\  they are quasi-topological cosmologies
whose characteristic functions carry the black-hole mass as a coupling. The
crossing criterion of the next section is untouched by any of this, since it
uses only continuity and the sign of the profile, and the prescription
spread of Section~\ref{sec:routeCimpl} now measures the distance between a
derived cosmology and two deformations of it. The statement is specific to the Lovelock and
quasi-topological class, in which one function governs the static and the cosmological sector
alike, as has long been known for Lovelock gravity \cite{DeruelleFarina1990}. In $f(R)$,
Horndeski or generic scalar--tensor theories the two sectors are not tied in this way, and
nothing of the kind is implied for them.

\section{The crossing criterion: monotone cores cannot cross the phantom divide}
\label{sec:criteria}

Before turning to the Hubble-radius routes, we establish a structural property
that holds across all three and that frames the confrontation with data. It
is most transparent in inverse form: instead of asking what cosmology a given
metric produces, ask which metric a prescribed history selects. Under each of
our maps the dark sector is a static profile $\rho_X(r)$ read along a
monotone curve $r=r(a)$, with the pressure fixed by continuity, $\dot\rho_{X}=-3H(1+\weff)\rho_{X}$. Since $\rho_{X}$ depends on time only through $r(a)$, the chain rule gives $\dd\ln\rho_{X}/\dd\ln a=(\dd\ln\rho_{X}/\dd\ln r)(\dd\ln r/\dd\ln a)$, and comparing the two expressions we obtain
\begin{equation}
1+\weff(a)=J(a)\,\big[1+w_X(r)\big]_{r=r(a)},\qquad
1+w_X(r)\equiv-\frac13\,\frac{\dd\ln\rho_X}{\dd\ln r},\qquad
J\equiv\frac{\dd\ln r}{\dd\ln a}.
\label{eq:darkcont}
\end{equation}
The profile differs between the routes. Route B (the self-conserved reading of route A/B, \cref{ssec:routeABfriedmann}) reads the local
Misner--Sharp density, $\rho_X=\rhoMS$, at $r=1/H$. Routes N and C read the
mean density inside the sphere, $\rho_X=\bar\rho_{\rm MS}=3m/4\pi r^{3}$, at
$r=a$ and $r=1/H$ respectively. The kinematic factor is $J=1$ for route N and
$J=\dd\ln H^{-1}/\dd\ln a=-\dot H/H^{2}$ for B and C. Its sign is not an extra assumption about the map. Differentiating the Friedmann
equation of either route (\cref{eq:routeABfriedmann,eq:routeCfriedmann} with $k=0$), and writing $\Omega_{X}$ for the
fraction of $H^{2}$ carried by the profile term,
\begin{equation}
J=\frac{\tfrac32\,\Omega_{m}+2\,\Omega_{r}}{1-\tfrac32\,(1+w_{X})\,\Omega_{X}}\,,
\label{eq:Jclosed}
\end{equation}
which is positive along the whole branch that descends from the matter era, vanishes only
at a de~Sitter fixed point ($\Omega_{m},\Omega_{r}\to0$), and diverges where the denominator
vanishes, which is the fold of \cref{ssec:Leffvalid}.

For the mean density the radial derivative is fixed by the local one through
the identity $r\,\bar\rho'_{\rm MS}=3(\rhoMS-\bar\rho_{\rm MS})$, so that
$1+w_X=1-\rhoMS/\bar\rho_{\rm MS}$, the relation already met in
\cref{eq:necidentity}. Denote by $a_\star$ a scale factor at which
$\weff(a_\star)=-1$ with $J\neq0$, and by $r_\star$ the corresponding radius.
\Cref{eq:darkcont} then answers at once which feature of the static profile
is being probed,
\begin{equation}
\weff(a_\star)=-1\iff
\begin{cases}
\rhoMS'(r_\star)=0 & \text{(route B),}\\[2pt]
\rhoMS(r_\star)=\bar\rho_{\rm MS}(r_\star) & \text{(routes N and C).}
\end{cases}
\label{eq:crossing}
\end{equation}
This is the central point of the argument. A value of the cosmological equation of
state, a statement about time evolution, is locked to a feature of the static
radial profile: a stationary point of the local density in route B, and a
point where the local density equals its own volume average in routes N and
C. (The exceptional case $J=0$ is the late-time de~Sitter attractor of
\cref{sec:routeABcosmo}, where $\weff\to-1$ is approached from above without
any feature of the profile and without a crossing.)

The direction of the crossing then fixes the \emph{type} of feature. A
Quintom-B crossing~\cite{Quintom,Quintom1}, i.e.\ $\weff<-1$ for $a<a_\star$
relaxing to $\weff>-1$ for $a>a_\star$, the direction favoured by DESI
DR2~\cite{DESIDR2}, requires $1+w_X$ to pass from negative to positive across
$r_\star$. In route B this means $\rhoMS$ first increases and then decreases,
a local \emph{maximum}, namely a density \emph{shell}. In routes N and C it
means $\rhoMS>\bar\rho_{\rm MS}$ before $r_\star$ and $\rhoMS<\bar\rho_{\rm MS}$
after, i.e.\ the mean density itself rises and then falls, which by the
identity above is possible only if the local density has risen somewhere
inside. Both conditions require a non-monotone $\rhoMS$.  For a shell that forms away from the centre the second is the stronger one, since the
shell must lift the average of everything inside it. A shell born at the centre has nothing
inside it to lift and the two thresholds coincide. This is the Jusufi--Singleton case: near the
origin $\rhoMS-\bar\rho_{\rm MS}\simeq M(M-15\ell)\,r^{2}/20\pi\ell^{6}$, so both conditions are met
as soon as $M/\ell>15$.

The consequence for regular black holes is immediate. The defining property
of a regular black hole with a de~Sitter core is that its Misner--Sharp
density decreases monotonically from a finite central value, $\rhoMS'(r)<0$
for all $r>0$. Then route B never meets the stationary point of
\cref{eq:crossing}, and routes N and C never meet the equality, because the
mean of a decreasing function always exceeds its endpoint value. In every
route $1+\weff>0$ wherever $J>0$: \emph{a monotone de~Sitter core can never
cross the phantom divide} and is quintessence-like throughout. The only way
to obtain a crossing is a non-monotone density with a maximum at finite
radius, which is a shell.

The result is prescription-independent in the sense that matters: it uses
only continuity, the definition of $\rhoMS$ and the sign of $J$, and the three
routes differ only in which profile (local or mean) is read and along which
curve. The interpretation is geometric. Indeed, a phantom crossing of the
cosmological equation of state is the same object as a density shell in the
metric. 
 The reconstruction of a CPL dark
sector~\cite{CPL,Linder2003},
$\rho_{\rm DE}(a)=\rho_{\rm DE,0}\,a^{-3(1+w_{0}+w_{a})}e^{-3w_{a}(1-a)}$, makes the
point concretely. In route B it is the local profile read at $r=1/H(a)$. For the DESI signs, $w_{a}<0$ and $1+w_{0}+w_{a}<0$, $\rho_{\rm DE}(a)$ is a power
times an exponential with a maximum at $a=(1+w_{0}+w_{a})/w_{a}$, and since $r(a)$ is monotone the profile
$\rhoMS(r)$ inherits a maximum at finite radius. In routes N and C it is the mean profile, and the identity
above returns the local one as $\rhoMS=-w_{X}\,\bar\rho_{\rm MS}$, with $w_{X}=w(a)$ in route N and $1+w_{X}=(1+w)/J$ in route C, so that the DESI
preference for $w<-1$ at $z\gtrsim0.5$ is the statement that the local density
exceeded its own volume average there. Either way the selected object is the profile
of a hairy or ``dirty'' black hole, not of a regular core. Among our six metrics only Jusufi--Singleton,
through the self-energy bump of its Misner--Sharp density, develops such a shell,
and only in the regime $M/\ell>15$ in which the $M^{2}$ self-energy term
dominates the Bardeen-like core term. (This is also the regime in which Jusufi and Singleton themselves regard the uncorrected $M^{2}$ term as unphysical and bound it by an interpolating function of $M/M_{\rm Pl}$~\cite{JusufiSingleton2025}. If the same replacement is made in the metric function, the self-energy never exceeds the core term and no shell forms. The crossing discussed here is a property of \cref{eq:mJS} as written.)  The threshold is exact: near the origin
$\rhoMS=3M/4\pi\ell^{3}+(M^{2}-15M\ell)\,r^{2}/8\pi\ell^{6}+\dots$, so the profile stops
decreasing at the centre when $M/\ell=15$ and the shell is born at the origin, moving
outward to $r\simeq0.16\,\ell$ at $M/\ell=16$ and towards $\ell/\sqrt2$ in the
pure-shell limit, and so tracks the regularisation length: under the inflationary normalisation of $\ell$
the crossing it would produce lies in the inflationary epoch, not at $z\lesssim2$,
while under the late-time normalisation of \cref{sec:routeABcosmo}, where
$\ell\sim H_{0}^{-1}$, the same shell would sit at the Hubble scale. In that regime
the source is no longer a regular core in the sense used here but a self-energy
shell, and the values of \cref{tab:cpl} stay on the core branch $M/\ell=1/2$. The dynamical dark energy favoured by DESI is
therefore, in this framework, a precise statement about metric structure: it
selects hairy black holes with a density shell, a class disjoint from the regular
cores at the centre.

Two comments are in order in view of recent work. First, the monotonicity
hypothesis of the criterion is the generic
energy-condition-respecting case. The weak energy condition, applied to the
anisotropic source through the radial continuity equation
$\rho'+\tfrac{3}{r}(\rho+\bar p)=0$ with $\bar p=(p_{r}+2p_{\perp})/3$, forces $\rho'(r)\le0$~\cite{Firouzjahi2026}, and the weak-energy-condition-satisfying regular black holes constructed in~\cite{CasadioKO2025} and analysed in~\cite{Easson2026} have just such monotone profiles. The
criterion therefore applies to the very class of physically admissible regular
cores and a density shell, which is the only escape, requires either a localised
violation of the weak energy condition or a source that is not a single regular
core. Second, the result should not be conflated with the source-side discussion
of Easson~\cite{Easson2026}, who notes (without proving a general criterion) that the
Bardeen source does not supply the persistent $w\le-1/3$ support needed for an
\emph{unbounded closed daughter}, since its static density falls as $r^{-5}$ and
its homogenised form is radiation-like. That is a statement about the magnitude
and decay of the support, aimed at the boundedness of a matched closed universe,
ours is a statement about the \emph{sign of the crossing}, that a monotone core
cannot reach $\weff=-1$ from below at all. The two are independent and mutually
consistent: the regular core is at once too weakly supporting to sustain
unbounded closed expansion and structurally unable to cross the phantom divide.

 The crossing criterion does coincide, however, with another part of the obstruction of \cite{Easson2026}. By the identity of
\cref{ssec:coresurvival},
$w_{\mathrm{eff}}<-1$ under the comoving identification means
$\rho_{\mathrm{MS}}>\bar\rho_{\mathrm{MS}}$, which is a violation of
the null energy condition in the matched FLRW region, so the crossing
criterion of this section coincides with the averaged-null-energy part of the obstruction
of~\cite{Easson2026}. Every construction in this paper escapes the obstruction in
one of two ways. A monotone core respects the averaged-null-energy condition and
gives up geodesic completeness instead
(\cref{ssec:coresurvival}), while a density shell can cross $w=-1$, but only by violating
the null-energy condition for some time.  The two formulations are now the two lines of \cref{eq:crossing}: the averaged-null-energy
condition of the matched region is the route-N line, and for Jusufi--Singleton  both conditions set in at $M/\ell=15$, a shell being needed in either case. We will
use one consequence of this in \cref{sec:routeAB}, i.e. a running vacuum built from a monotone core can
never itself endanger the averaged null energy condition, since no such core
produces a phantom component.

 We should finally be clear about the logical status of the criterion. It constrains
 the reconstruction map and not dark energy as such: if the dark sector is not the
 Misner--Sharp density of a static profile evaluated along a monotone $r(a)$, the
 criterion has nothing to say. Within the maps considered in this paper, however,
 it is exhaustive.

\section{Route A/B: the running vacuum}
\label{sec:routeAB}

\subsection{Why $\Leff$ is a legitimate vacuum: the equation of state}
\label{ssec:Leffvalid}
That $\Leff$ in \cref{eq:Leffdef} deserves the name of a vacuum is not just a manner of speaking, it follows from the equation of state of the core and we will now show how.
For any metric of the form \eqref{eq:ssmetric} the Einstein tensor is diagonal,
and the condition $g_{tt}\,g_{rr}=-1$, equivalently the absence of an
independent function multiplying $\dd t^{2}$ beyond $f$, which holds for every metric written in the form (\ref{eq:ssmetric}), hence for all six of the ones chosen in this paper, forces the time--time and radial--radial components to
coincide, $G^{t}{}_{t}=G^{r}{}_{r}$. Reading the stress tensor off
$G^{\mu}{}_{\nu}=8\pi G\,T^{\mu}{}_{\nu}$ with
$T^{\mu}{}_{\nu}=\mathrm{diag}(-\rho,p_{r},p_{\perp},p_{\perp})$, this identity is
\begin{equation}
  p_{r}(r)=-\rho(r)=-\rhoMS(r),
  \label{eq:pr}
\end{equation}
\emph{exactly}, not as an approximation, where the second equality uses the
Misner--Sharp identification $\rho=m'/4\pi r^{2}$~\cite{MisnerSharp1964}. The
source that supports a regular black hole of this form is therefore necessarily
anisotropic, with radial equation of state $w_{r}=-1$ at every radius. This is a
generic feature of regular cores and is the precise sense in which the interior
is vacuum-like~\cite{Dymnikova1992,Dymnikova2002,AnsoldiReview2008}. At the
centre regularity additionally forces isotropy, $p_{r}=p_{\perp}$, so that
$p_{r}=p_{\perp}=-\rhoMS=-\Lcore/8\pi G$ and the core is a cosmological constant
in the strict sense, away from the centre the tangential pressure differs (the anisotropy
$p_{\perp}-p_{r}$ is what stores the profile's radial dependence) and $\Leff(r)$
is a \emph{position-dependent} vacuum. Promoting this to cosmology, the density
\begin{equation}
  \rho_{\Lambda}(a)=\frac{\Leff(\rH)}{8\pi G},\qquad \rH=H^{-1},
\end{equation}
is a running vacuum: its equation of state is exactly $-1$ in the adiabatic
(core) limit and departs from $-1$ as the profile runs. We denote it $\Ltil$ to
emphasise that it is a dynamical term, not a true cosmological constant, the
identification $\rho_{\Lambda}=\Ltil/8\pi G$ is exact only where $\Leff$ is flat.

This places route A/B in the well-developed framework of running-vacuum
cosmologies. Position or scale dependent vacuum energy was proposed by
Dymnikova in the black-hole context~\cite{Dymnikova2002} and developed
 as a renormalisation-group running of $\Lambda$ by Shapiro and
Sol\`a~\cite{ShapiroSola2002} and by Sol\`a and collaborators~\cite{Sola2013,SolaGomezPeracaula2015}.
The bookkeeping of any energy exchange with matter follows the
interacting-dark-energy literature~\cite{Wang2016}. Our contribution is to fix
the running not by a phenomenological ansatz but by a specific regular-black-hole
profile evaluated at the Hubble radius.

This route is, moreover, the kind of construction the minimal-daughter
obstruction calls for. Easson~\cite{Easson2026} shows that escaping the
boundedness of the minimal closed daughter requires ``an additional smooth bulk
component whose density redshifts no faster than $A^{-2}$,'' of which a positive
vacuum-energy component is the simplest example, and which lies by construction
outside the minimal no-shell matching. The running vacuum $\tilde\Lambda_{\mathrm{eff}}$ added in
\cref{eq:routeABfriedmann} is a component of just this kind throughout the past and
across the entire fitted range, where the Hubble radius has not yet run far
beyond the knee of the profile. Whether it remains one as $a\to\infty$ (the
clause asks for persistent support, strong enough to forestall the turnaround
of a closed daughter) is decided by the fixed-point condition of
\cref{sec:routeABcosmo}:  the support is permanent when the pure-vacuum relation
$H^{2}=\tilde\Lambda_{\mathrm{eff}}(1/H)/3$ admits a positive root, in which
case the expansion settles onto a late-time de~Sitter attractor and the
criterion is met trivially. Below that threshold two things can happen, and
which one depends on the regularisation length. Either the vacuum eventually
redshifts far faster than $a^{-2}$ (as $a^{-9}$ for Hayward in a matter
background) and the support, while real, is transient, or the implicit
equation~\eqref{eq:routeABfriedmann} reaches a fold, $2H=\dd F/\dd H$ with
$F\equiv\tilde\Lambda_{\mathrm{eff}}(1/H)/3$, beyond which it has no real
solution. At the fold the scale factor and the Hubble rate are finite while
$\dot H\to-\infty$, a sudden singularity of the type classified by
Barrow~\cite{Barrow2004}. The fold is a property of the implicit map
$\rho_{\Lambda}=F(H)$ and marks where the Hubble-radius identification stops
defining a cosmology, which is consistent with its status in
\cref{ssec:ABCnotderived} as a relation whose couplings carry the black-hole mass.
We return to the three outcomes, and to which of them the fiducial scales of
\cref{tab:cpl} realise, in \cref{sec:routeABcosmo}. The escape from boundedness
is therefore conditional rather than automatic, and we state it as such. Routes A/B and C should nonetheless be read not as competitors to the
minimal construction but as the kind of additional structure it shows to be
necessary, with the fixed-point condition marking where that structure is
permanent.

\subsection{Friedmann equations}
\label{ssec:routeABfriedmann}

We promote the profile through $r\mapsto\rH=H^{-1}$ and add it to ordinary matter
and radiation. The first Friedmann equation is
\begin{equation}
  H^{2}=\frac{8\pi G}{3}\Big(\rho_{m}+\rho_{r}+\frac{\Leff(H^{-1})}{8\pi G}\Big)-\frac{k}{a^{2}},
  \label{eq:routeABfriedmann}
\end{equation}
implicit in $H$ because the profile is evaluated at $r=1/H$ on the right-hand
side. There are two ways to book the time-dependence of the vacuum,
and they share this equation. In \emph{route B} one treats $(\rho_{\Lambda},p_{\Lambda})$
as a self-conserved fluid, and continuity then fixes
\begin{equation}
  \weff(a)=-1-\frac{1}{3}\frac{\dd\ln\Leff}{\dd\ln a}
          =-1-\frac{1}{3}\frac{\dd\ln\Leff}{\dd\ln r}\Big|_{r=1/H}\frac{\dd\ln H^{-1}}{\dd\ln a},
  \label{eq:weffAB}
\end{equation}
where the second factor is the kinematic Jacobian, non-negative along the physical branch (\cref{eq:Jclosed}). This is the route-B instance of the general split of
\cref{eq:darkcont}, in which the anisotropic source of the static metric is
isotropised at the background level by reading a single profile along $r(a)$, the
tangential pressure entering only through the continuity equation. The statement is exact on the static side. With $p_{r}=-\rhoMS$, radial conservation gives $p_{\perp}=-\rhoMS-\tfrac12\,r\rhoMS'$, so the angle-averaged pressure $\bar p\equiv(p_{r}+2p_{\perp})/3$ obeys
\begin{equation}
\frac{\bar p}{\rhoMS}=-1-\frac13\,\frac{\dd\ln\rhoMS}{\dd\ln r}=w_{X}(r),
\label{eq:isotrop}
\end{equation}
and the pressure that continuity assigns to the cosmological fluid is the isotropic average of the principal pressures of the static source, rescaled by the kinematic factor of \cref{eq:Jclosed}: $1+\weff=J\,(1+\bar p/\rhoMS)$ at $r=1/H$. The homogeneous vacuum is thus the angular average of the anisotropic source and not an independent postulate about it. In \emph{route A} one insists $w_{\Lambda}\equiv-1$ exactly and
routes the running into a vacuum-to-matter transfer
$Q=-\dot\rho_{\Lambda}\propto\dot H$, with $\dot\rho_{m}+3H\rho_{m}=Q$, so that the vacuum is then de~Sitter at each instant.  The two bookkeepings share the first Friedmann equation but not the
expansion history, because in route A the pressure of the dark sector is
$-\rho_{\Lambda}$ while in route B it is $\weff\rho_{\Lambda}$, and the two
histories coincide only where $\Leff$ is flat. At the scales that place $w_{0}$
in the DESI band, the transfer of route A creates a sizeable fraction of the
present matter density after $z\sim3$, which is incompatible with the CMB value
of $\omega_{m}$. We will therefore use the self-conserved reading throughout,
and the values of \cref{tab:cpl} refer to it. We keep the label ``A/B'' for the
route since the two readings share the source function, but every number in
this paper is route B. We
stress, in line with the crossing criterion, that neither bookkeeping produces a phantom
component from a monotone decaying core: $\Leff$ decreases once the Hubble radius
exceeds the core scale, so the metric factor in \cref{eq:weffAB} is negative and
$\weff>-1$ throughout. This is the route-B line of \cref{eq:crossing}, and it holds with
the kinematic factor $J$ included, since $J\geq0$ for the self-consistent expansion and
$J=0$ only at the fixed point of \cref{sec:routeABcosmo}. The Friedmann pair is completed in the standard way: the acceleration equation
$\ddot{a}/a=-\frac{4\pi G}{3}\sum_{i}(\rho_{i}+3p_{i})$ holds with $p_{\Lambda}=\weff\,\rho_{\Lambda}$
in route B (route A would instead carry $p_{\Lambda}=-\rho_{\Lambda}$ and the transfer $Q$, and a different history), so no input enters beyond
\cref{eq:routeABfriedmann} and continuity.

For Hayward the explicit source is
\begin{equation}
  \frac{\Leff(H^{-1})}{3}=\frac{4G^{2}M^{2}\ell^{2}H^{6}}{(1+2GM\ell^{2}H^{3})^{2}},
\end{equation}
and the equation $H^{2}=\Omega_{m0}a^{-3}+\Omega_{r0}a^{-4}+\Leff(H^{-1})/3$ admits
an exact inversion $a(H)$: although it is transcendental in $H$ (the profile is
evaluated at $r=1/H$), it contains the scale factor only through the explicit
$\Omega_{m0}a^{-3}$ and $\Omega_{r0}a^{-4}$ terms, so at fixed $H$ it is algebraic
in $a$, and dropping radiation it is linear in $a^{-3}$,
$a(H)=[\Omega_{m0}/(H^{2}-\Leff(H^{-1})/3)]^{1/3}$. The time then follows from a rational quadrature.

 A remark on the interpretation is worthwhile here. At the late-time normalisation the
 fitted scales are $\ell \sim H_{0}^{-1}$ with $M\sim \ell/2$, so the metric no longer describes
 an astrophysical object. The regular black-hole solution acts as a generating profile for the
 running law $\tilde{\Lambda}(H)$, its two parameters inherited from the black-hole family
 but no longer tied to a localised mass. The black-hole reading is exact only in the inflationary normalisation, whereas at late times the construction is a phenomenological running-vacuum model
 whose functional form is what confronts the data. After the amplitude normalisation below, four of the six
 geometries reduce to one-parameter families in a single length, and only Jusufi--Singleton and
 Bonanno--Reuter retain the ratio $M/\ell$ as a genuine second shape parameter (\cref{sec:routeABcosmo}).

\section{Route A/B: core regime, late-time vacuum, and DESI}
\label{sec:routeABcosmo}

 The route A/B history has two regimes joined by the running of $\Leff$. The first is a pure-vacuum limit: we recall from \cref{ssec:coresurvival} that with independently conserved matter and radiation the vacuum term is subdominant toward the past, so that the core regime described next is not reached along the standard thermal history. Deep in
the core ($H^{-1}\ll\ell$) the vacuum saturates to $\Lcore$ and, \emph{when it
dominates the matter content}, drives a de~Sitter phase $H\simeq1/\ell$. This is the inflationary regime, realised for the scale $\ell\sim10^{4}\ell_{\rm Pl}$ (well above the Planck length) fixed by the inflationary amplitude, and
self-consistent only above a threshold mass. For Hayward the pure-vacuum fixed
point $H^{2}=\Leff(H^{-1})/3$ reduces, on taking the positive square root, to
the cubic $2GM\ell^{2}H^{3}-2GM\ell H^{2}+1=0$, whose discriminant is
proportional to $8M-27\ell$ (in $G=1$ units): a positive constant-$H$ solution
exists only for $M\ge27\ell/8$, with the marginal root at $H_{\ast}=2/(3\ell)$.
The same cubic plays two roles. With the inflationary normalisation its relevant root is the self-consistent plateau just described, whereas with the late-time normalisation of the dark-energy comparison below, the
same equation, its amplitude rescaled accordingly, decides the fate of the
expansion. When the corresponding discriminant condition holds, the
decreasing Hubble rate is bounded below by the largest root and settles onto
it, so the expansion ends in a second, low-scale de~Sitter phase, the
permanent support anticipated in \cref{ssec:Leffvalid}. When it fails, the
outcome is decided by whether the branch meets the fold of
\cref{ssec:Leffvalid} before the vacuum has decayed. At the fiducial scales of
\cref{tab:cpl} the three outcomes are all realised: Dymnikova and NSS settle
onto attractors at $H_{\infty}\simeq0.77H_{0}$ and $0.80H_{0}$, Bardeen decays
(though it lies within one percent of the fold threshold), and Hayward,
Jusufi--Singleton and Bonanno--Reuter meet a fold at $a\simeq1.36$, $1.72$
and $2.28$, i.e.\ roughly $5$, $10$ and $15\,$Gyr from now. None of this
touches the fitted range, which ends at $a=1$, but it does mean that the late-time fate
of a running regular core is quite sensitive to $\ell$ and is not the
uniform slow decay one might expect. A single condition on $M/\ell$ thus
controls both self-consistent inflation and the persistence of the late-time
vacuum, and a second one, the fold, separates transient support from a
sudden end. The fold and the position in the $(w_{0},w_{a})$ plane are not independent. By \cref{eq:Jclosed} a single combination, $\tfrac32(1+w_{X})\Omega_{X}$, controls both: it enhances $J$, and with it $1+\weff$ today, and it is what reaches unity at the fold. For Hayward and Bardeen at the fiducial scales it equals $0.40$ at $a=1$, so that $J(a{=}1)\simeq0.79$ against $0.47$ for $\Lambda$CDM with the same $\Omega_{m0}$ (and $0.54$ for Dymnikova), i.e.\ the distance of $w_{0}$ from $-1$ in \cref{tab:cpl} is amplified by a factor of about $1.7$ through the implicit Hubble-radius map. The metrics that reach the DESI $w_{0}$ band are those that later meet a fold or, for Bardeen, lie within one percent of it, while the two attractor cases stay close to $\Lambda$CDM. The agreement at the level of $w_{0}$ should be read with this amplification in mind. At late times the profile has decayed and
$\Ltil$ acts as a slowly varying dark energy on top of matter and radiation. The
two regimes require different scales (a manifestation of the cosmological
constant problem, in which the scale that regularises the centre and the scale
that controls late-time acceleration differ by some sixty orders of magnitude),
so route A/B describes inflation and dark energy as separate sectors of the same
construction, not as a unification.

Fixing the amplitude by $\Omega_{\Lambda}(a{=}1)=0.685$ absorbs the overall
constant of $\Leff$. Explicitly, in all numerical work the dark-sector term of
\cref{eq:routeABfriedmann} and of \cref{eq:routeCfriedmann} is taken as
\begin{equation}
  \frac{8\pi G}{3}\,\rho_{X}\;\longrightarrow\;
  \Omega_{\Lambda0}\,H_{0}^{2}\,\frac{\rho_{X}(1/H)}{\rho_{X}(1/H_{0})},
  \qquad \Omega_{\Lambda0}=1-\Omega_{m0}-\Omega_{r0},
  \label{eq:normsource}
\end{equation}
with $\rho_{X}=\rhoMS$ for route B and $\rho_{X}=\bar\rho_{\rm MS}$ for route C, so that the
value $M=\ell/2$ quoted in the tables fixes the shape of the profile and not its amplitude.
Taken literally, the Hayward source of \cref{ssec:routeABfriedmann} with $M=\ell/2$ could not
reach $\Omega_{\Lambda0}$ for any $\ell$. This normalisation does not fix $\ell$. For Hayward, Bardeen, Dymnikova and
NSS the mass and the regularisation length enter the normalised profile only
through one degenerate length combination (for Hayward $(2M\ell^{2})^{1/3}$, for
Dymnikova $r_{0}$, for Bardeen $g$ and for NSS $\sqrt\theta$), so each of these
reduces to a one-parameter family and the value of $\ell$ quoted in
\cref{tab:cpl} is a fiducial choice rather than a derived quantity. The fiducial values are of the order of the Hubble radius and were fixed once, as a single illustrative set that serves the present-day indicators of \cref{tab:routeNindicators}, the parameters of \cref{tab:cpl} and the route C values of \cref{eq:routeCnumbers} alike. No fit is involved. Only
Jusufi--Singleton, whose core term is linear in $M$ while its self-energy term is
quadratic, and Bonanno--Reuter, whose running of $G$ ties $M$ to the scale of the
profile, retain $M/\ell$ as a genuine second shape parameter. With the amplitude
fixed in this way and the fiducial scales of \cref{tab:cpl}, and
integrating the late-time dark sector, we obtain the CPL parameters~\cite{CPL,Linder2003} of
each metric by fitting its effective equation of state to the linear form
$w(a)=w_{0}+w_{a}(1-a)$, in the least-squares sense over $a\in[0.55,1]$. We note that the fit depends only mildly on this window (over $a\in[0.3,1]$ Hayward gives $(-0.706,-0.422)$ and Bardeen $(-0.677,-0.364)$), whereas the DESI constraints derive from
a full likelihood over data reaching $z\simeq 2.3$, so the comparison is indicative not statistical. The cosmological
parameters are $H_{0}=1$, $\Omega_{m0}=0.315$, $\Omega_{r0}=9.16\times10^{-5}$.
\Cref{tab:cpl} reports the result against the three DESI DR2 dataset
combinations.

\begin{table}[t]
\centering
\renewcommand{\arraystretch}{1.3}
\begin{tabular}{|l|c|c|c|}
\toprule
Dataset / Model & $\ell$ ($H_{0}^{-1}$) & $w_{0}$ & $w_{a}$ \\
\midrule
\multicolumn{4}{|l|}{\textit{DESI DR2 $w_{0}w_{a}$CDM constraints (68\% CL)~\cite{DESIDR2}}} \\
\quad DESI+CMB+Pantheon$^{+}$ & --- & $-0.838\pm0.055$ & $-0.62^{+0.22}_{-0.19}$ \\
\quad DESI+CMB+Union3        & --- & $-0.667\pm0.088$ & $-1.09^{+0.31}_{-0.27}$ \\
\quad DESI+CMB+DESY5         & --- & $-0.752\pm0.057$ & $-0.86^{+0.23}_{-0.20}$ \\
\midrule
\multicolumn{4}{|l|}{\textit{Regular-black-hole geometries}} \\
\quad Hayward          & $1.6$ & $-0.700$ & $-0.449$ \\
\quad Bardeen          & $1.8$ & $-0.687$ & $-0.311$ \\
\quad Dymnikova        & $2.0$ & $-0.926$ & $-0.082$ \\
\quad NSS              & $2.0$ & $-0.953$ & $-0.017$ \\
\quad Jusufi--Singleton & $1.8$ & $-0.702$ & $-0.285$ \\
\quad Bonanno--Reuter  & $1.9$ & $-0.809$ & $-0.161$ \\
\midrule
\quad $\Lambda$CDM     & ---   & $-1.000$ & $\phantom{-}0.000$ \\
\bottomrule
\end{tabular}
\caption{ CPL parameters $(w_{0},w_{a})$ for the six geometries in route A/B
(amplitude-normalised, continuity-fixed sign, $M=\ell/2$ with $\ell$ each
metric's own regularisation scale), against the DESI
DR2 $w_{0}w_{a}$CDM constraints. The quoted $\ell$ are fiducial choices: for the
four one-parameter geometries (Hayward, Bardeen, Dymnikova, NSS) the shape of
$w(a)$ depends on $\ell$ alone once the amplitude is normalised, and Jusufi--Singleton
is shown on its core branch $M/\ell=1/2$. Every model shown is thawing quintessence. The rational cores reach the DESI $w_{0}$ band at roughly half the $|w_{a}|$, and NSS is observationally degenerate with Dymnikova.}
\label{tab:cpl}
\end{table}

Three observations follow. First, every viable model is thawing quintessence,
$w_{0}>-1$ relaxing from a more negative past value, consistent with
the barrier: the monotone cores cannot cross $-1$. Second, the rational cores
(Hayward, Bardeen) reach the DESI $w_{0}$ band but at roughly half the required
$|w_{a}|$, while the Gaussian cores (Dymnikova, NSS) sit close to $\Lambda$CDM with
very small $|w_{a}|$: the decay law of the profile, not the universal core,
controls the position in the plane. Because $\ell$ is fiducial, this second
observation must be read as a statement about the trajectory each geometry traces
in the $(w_{0},w_{a})$ plane as $\ell$ runs, and we have checked that the four one-parameter cores leave $\Lambda$CDM along nearly a single curve that lies above the
DESI degeneracy direction, at roughly half its $|w_{a}|$ for a given $w_{0}$. The particular entries of \cref{tab:cpl} are points on that curve, and a fit of $\ell$
to the DESI distances themselves is left to future work. Third, none of the
monotone cores reproduces the central DESI preference for a phantom crossing,
which by the shell argument above is structurally unavailable to them. In sum, route A/B yields viable thawing quintessence, consistent with the data
at the $w_{0}$ level, but does not
realise the specific crossing that motivates the dynamical-dark-energy
interpretation of DESI.

The crossing therefore works as a test and not as a failure of the model, and both possible outcomes are interesting.
If the crossing consolidates, the whole class of monotone regular cores is ruled
out as a description of the dark sector and the data points towards the density-shell geometries which,
as we saw in  \cref{sec:criteria}, violate the energy conditions. If it fades away (the DESI DR2 Lyman-$\alpha$
measurements are consistent with $\Lambda$CDM~\cite{DESIDR2Lya}, and how stable the signal is across data
releases is still being debated~\cite{OColgain2025}), then a thawing quintessence which reaches the $w_{0}$ band with
roughly half the CPL $|w_{a}|$ is exactly the region occupied by these geometries. In either case the framework
can be falsified, since it forbids $w<-1$ at every epoch and for every route as long as the
source is a monotone core, and permits it only where the source develops a density shell,
which among our metrics happens only for Jusufi--Singleton in the self-energy-dominated
regime $M/\ell>15$ of \cref{sec:criteria}, outside the range of \cref{tab:cpl}.

The comparison just made lives in the $(w_{0},w_{a})$ plane, which is the plane
in which the data are reported but not the one in which our crossing criterion is most transparent, and it is worthwhile to explain why. For a spatially flat universe of
dust plus a dark sector of equation of state $w(z)$, write
$E^{2}(z)=\Omega_{m0}(1+z)^{3}+(1-\Omega_{m0})f(z)$ with the dark-sector density
$f$ fixed by continuity, $\mathrm{d}\ln f/\mathrm{d}z=3(1+w)/(1+z)$, and let
$\Omega_{\rm DE}(z)=(1-\Omega_{m0})f/E^{2}$ be its running fraction. The
deceleration parameter is then the familiar
\begin{equation}
  q(z)=-1+(1+z)\frac{E'(z)}{E(z)}
      =\tfrac12\Big[1+3\,w(z)\,\Omega_{\rm DE}(z)\Big],
  \label{eq:qtwofluid}
\end{equation}
and one further derivative through $j=q(2q+1)+(1+z)\,q'$~\cite{Visser2004},
using the continuity relation to eliminate $f'$, gives
\begin{equation}
  j(z)-1=\frac{3}{2}\,\Omega_{\rm DE}(z)
         \Big[\,3\,w(1+w)+(1+z)\,\frac{\mathrm{d}w}{\mathrm{d}z}\Big].
  \label{eq:jerkid}
\end{equation}
The two terms inside the bracket carry different information, and at the phantom
divide only one of them survives: $w=-1$ annihilates $3w(1+w)$ identically, so at
a crossing scale $z_{\star}$ the whole departure of the jerk from unity is the
slope of the equation of state,
\begin{equation}
  j(z_{\star})-1=\frac{3}{2}\,\Omega_{\rm DE}(z_{\star})\,(1+z_{\star})\,
                 w'(z_{\star}),
  \label{eq:jerkcross}
\end{equation}
which is negative in the Quintom-B direction, where $w$ increases with $a$. The
deceleration parameter is by contrast entirely featureless there: at $z_{\star}$
it takes the value $q=\tfrac12[1-3\Omega_{\rm DE}(z_{\star})]$, which marks
nothing. A crossing of the phantom divide is therefore invisible at the level of
$q$ and lives wholly in the jerk.

This is the observational counterpart of the criterion of \cref{sec:criteria}.
Because a monotone core keeps $1+\weff>0$ throughout, the routes never reach
$w=-1$, and so cannot generate the negative jerk excursion that
\cref{eq:jerkcross} attaches to a crossing. Hence, the most direct discriminating observable is $j(z)$ rather than the fitted pair $(w_{0},w_{a})$, which averages
the history over the fitting window and reports the crossing only indirectly. We
note that \cref{eq:jerkid} applies to the two-component split used in routes A/B
and C, and should not be conflated with the single-fluid jerk of route N in
\cref{ssec:cosmography}, where the departures of $j_{0}$ from unity recorded in
\cref{tab:routeNindicators} are transition effects of the running core rather
than crossing effects. The practical caveat is that kinematic determinations of
$j_{0}$ are far weaker than those of $q_{0}$, so this is at present a
discriminator in principle rather than a live constraint.

A cruder cosmographic check bears on how much weight the $(w_{0},w_{a})$ target
should carry in the first place. Present-day acceleration in the CPL
parametrisation requires $q_{0}<0$, that is
$|w_{0}|>[3(1-\Omega_{m0})]^{-1}$, a threshold lying between $0.48$ and $0.52$
for $\Omega_{m0}$ in the range $0.31$--$0.36$. The BAO$+$CMB central value
$w_{0}=-0.42$~\cite{DESIDR2} does not meet it, so the corresponding $q_{0}$ is
positive, against model-independent determinations, which consistently give $q_{0}<0$~\cite{MehrabiRezaei2021}. This tension was pointed out for the DR1 fit
in Ref.~\cite{BaticBMN2024} and is not removed by DR2. We do not read it as
evidence against the data, but as a reason to treat the CPL plane as an
indicative target rather than a fit, and hence not to regard a construction that
undershoots its central $|w_{a}|$ as thereby excluded.

\section{Route C: the quasi-local construction}
\label{sec:routeC}

The third route replaces the local density by a quasi-local mass. For spatially
flat FLRW the apparent horizon coincides with the Hubble sphere,
$R_{\rm AH}=1/H$~\cite{Faraoni2015}, and the gravitating energy relevant at each
epoch is the Misner--Sharp--Hernandez mass $m(R_{\rm AH})$ enclosed within it.
Spreading this mass over the Hubble volume defines a mean cosmological density,
\begin{equation}
  \rho_{\rm eff}(a)=\frac{3\,m(1/H)}{4\pi H^{-3}}=\frac{3H^{3}}{4\pi}\,m(1/H),
  \label{eq:rhoCgen}
\end{equation}
and the first Friedmann equation reads
\begin{equation}
  H^{2}=\frac{8\pi G}{3}\Big(\rho_{m}+\rho_{r}+\frac{3H^{3}}{4\pi}m(1/H)\Big)-\frac{k}{a^{2}}.
  \label{eq:routeCfriedmann}
\end{equation}
From here on we set $k=0$ in this route, since for $k\neq0$ the apparent horizon lies at $(H^{2}+k/a^{2})^{-1/2}$ and not at $1/H$.
The construction assumes the static metric satisfies $g_{tt}\,g_{rr}=-1$, so that
an FLRW source can be read off directly. This holds for all six metrics of \cref{sec:metrics}, since each is written in the
form~\eqref{eq:ssmetric}. Route C is nonetheless not the cosmological sector of the
theory whose vacuum is the black hole, its characteristic function carries the mass
as a coupling (\cref{ssec:ABCnotderived}), and we therefore read it as a measure of
prescription sensitivity rather than as a derived theory.
The source term that route C adds to $\tfrac{8\pi G}{3}(\rho_{m}+\rho_{r})$ is
$2GH^{3}m(1/H)$, to be compared with the local $\Leff(1/H)/3$ of route A/B. For the reference case,
\begin{equation}
  \text{(Hayward)}\quad 2GH^{3}m(1/H)=\frac{2GMH^{3}}{1+2GM\ell^{2}H^{3}},
\end{equation}
and the per-metric forms are collected in \cref{tab:routeC}. As in route A/B the
equation admits an exact inversion $a(H)$, and the time integral is elementary for
the rational cores and numerical for the Gaussian ones.

\begin{table}[t]
\centering
\renewcommand{\arraystretch}{1.6}
\begin{tabular}{|l>{$}l<{$}c|}
\toprule
Metric & \text{Route C source } 2GH^{3}m(1/H) & late-time decay \\
\midrule
Hayward    & \dfrac{2GMH^{3}}{1+2GM\ell^{2}H^{3}}            & $\sim H^{3}$ \\
Bardeen    & \dfrac{2GMH^{3}}{(1+g^{2}H^{2})^{3/2}}          & $\sim H^{3}$ \\
Dymnikova  & 2GMH^{3}\big(1-e^{-H^{-3}/r_{0}^{3}}\big)       & $\sim H^{3}$ \\
NSS        & 2GMH^{3}\,\mathrm{erf}\!\big(\tfrac{1}{2H\sqrt\theta}\big)-\tfrac{2GMH^{2}}{\sqrt{\pi\theta}}e^{-1/4\theta H^{2}} & $\sim H^{3}$ \\
Bonanno--Reuter & \dfrac{2G_{0}MH^{3}}{1+\tilde\omega G_{0}H^{2}(1+\gamma G_{0}MH)} & $\sim H^{3}$ \\
Jusufi--Singleton & 2G\,H^{3}\,m(1/H)\ \text{(\cref{eq:mJS})} & $\sim H^{3}$ \\
\bottomrule
\end{tabular}
\caption{Route C source terms. Every source decays as $H^{3}$ at late times,
because the enclosed mass saturates at the ADM value $m\to M$ ($\mathcal{M}$ for Jusufi--Singleton) regardless of the
metric's asymptotic profile, the universal feature that places route C
considerably closer to $\Lambda$CDM than route A/B (whose local source decays as
$H^{4}$--$H^{6}$).}
\label{tab:routeC}
\end{table}

\section{Route C: cosmological implications}
\label{sec:routeCimpl}

The defining feature of route C is universal: as $H\to0$ the enclosed mass
saturates at the ADM value, $m(1/H)\to M$ ($\mathcal{M}$ of \cref{eq:mJS} for Jusufi--Singleton), so every source term decays as
$2GMH^{3}$ irrespective of the metric. This volume-averaged saturation makes the
quasi-local density evolve more slowly than the local Misner--Sharp density of
route A/B, and the cosmological consequence is systematic: route C sits closer to
$\Lambda$CDM than route A/B at every redshift. Integrating the late-time sector
over the observational window gives, at the fiducial scales,
\begin{equation}
  (w_{0},w_{a})_{\rm C}=
  \begin{cases}
    (-0.875,\,-0.134) & \text{Hayward, }\ell=1.6,\\
    (-0.843,\,-0.082) & \text{Bardeen, }\ell=1.8,\\
    (-0.966,\,-0.033) & \text{Dymnikova, }\ell=2.0,
  \end{cases}
  \label{eq:routeCnumbers}
\end{equation}
to be compared with the route A/B values $(-0.700,-0.449)$, $(-0.687,-0.311)$ and
$(-0.926,-0.082)$ of \cref{tab:cpl}. The shift toward $w_{0}=-1$ is uniform, and
the prescription spread, $\Delta w_{0}\simeq0.17$ (Hayward), $0.16$ (Bardeen),
$0.04$ (Dymnikova), is a measure of the prescription dependence of the construction. For the rational cores it exceeds the DESI DR2 $w_{0}$
uncertainty of $\sim0.06$--$0.09$. The asymptotic equation of state also differs:
continuity applied to the saturated source gives $\weff\to-1+\dd\ln H^{-1}/\dd\ln a$,
which is $+\tfrac12$ in a matter-dominated background rather than the dust value
$0$ of the naive expectation. That limit requires $H\to0$, and at the fiducial scales of \cref{eq:routeCnumbers} it is
never reached. With the normalisation of \cref{eq:normsource} the pure-vacuum relation of
route C has a root below $H_{0}$, and the expansion settles onto a de~Sitter attractor,
$H_{\infty}\simeq0.71H_{0}$ for Hayward, $0.70H_{0}$ for Bardeen and $0.80H_{0}$ for Dymnikova, so that
$\weff\to-1$ in the future. The $+\tfrac12$ limit can only be reached at smaller $\ell$, where the
relation has no such root. At the scale where route B meets a fold for Hayward, the volume average of route C
softens the source enough to replace the fold by a smooth attractor. Route C is not
free of folds either: for Hayward one opens for $1.3\lesssim\ell H_{0}\lesssim1.4$, between the decaying
and the attractor regimes. By $z=0$ the equation of
state is still drifting upward from $-1$. We stress that route C does not cross $w=-1$ either. Its source is the mean density inside the Hubble sphere, so it obeys
the routes-N-and-C line of \cref{eq:crossing}, and a monotone core never lets the local
density reach its own average.

The main result of the route C analysis is thus twofold.  The agreement on the core is a statement about the source term, which saturates to
$\Lcore/3$ in every route as its argument shrinks, and in this sense the core is a
property of the geometry and not of the prescription. Whether the expansion ever sits
on that value is a different matter: it does in route N, where nothing else is present
and where, for the metrics of \cref{ssec:hexists}, the equation is that of a covariant
theory, and it does not in routes A/B and C, where independently conserved matter and
radiation dominate toward the past  (\cref{fig:coresurvival}). The disagreement at late times ($H^{3}$ decay for route C versus $H^{4}$--$H^{6}$ for route A/B) is a real prescription dependence, and it limits how precisely any single prescription can predict the dark-energy parameters. A
measurement of $(w_{0},w_{a})$ at the few-percent level does not select a metric
without first selecting a route.

\section{Conclusions and outlook}
\label{sec:classification}

Let us close by summarising the results, since the value of treating three
routes and six metrics in one place is the pattern that emerges across them.

Route N is the most informative about the early universe. Its Friedmann equation
coincides with the minimal no-shell matched daughter derived rigorously
in~\cite{Easson2026} and, for the metrics of \cref{ssec:hexists}, with the FLRW sector of the quasi-topological theory (\cref{sec:covariant}). The static interior itself is Kantowski--Sachs~\cite{CasadioKO2025} and is not what route N describes (\cref{sec:routeN}). What the present treatment adds, on top of that
now-established equation, is the content that follows from carrying it across six metrics, namely the critical-exponent classification of which metrics regularise
which barotropic cosmology, the fully analytic solution for Hayward and the
analytic-status grid for the rest, and the single-function ($\mu$) cosmography of
\cref{eq:cosmography}. Read as a standalone homogeneous model, route N is  curvature-regular, and it identifies a \emph{primordial}, inflationary
acceleration with a graceful exit to dust. Consistent with its inflationary
character, and with the obstruction theorem of~\cite{Easson2026}, its de~Sitter
past is past null incomplete in the same way standard inflation is. It offers no
late-time dark energy, and its completion by a bare $\Lambda$ fails because of the scale hierarchy. Routes A/B and C are informative about late times, and supply the kind of additional structure the
minimal-daughter obstruction shows to be necessary (permanently so when the pure-vacuum fixed point of \cref{sec:routeABcosmo} exists, and otherwise either transiently or up to a fold at finite $a$). Both yield thawing quintessence consistent with the DESI $w_{0}$ band, A/B from
the local Misner--Sharp density and C from the quasi-local enclosed mass, with C
systematically closer to $\Lambda$CDM. The three routes agree on the core value the geometry supplies, though only route N ever sits on it, and differ at late times by a prescription-dependent spread that measures
how much the late-time predictions owe to the choice of map rather than to the metric. That agreement is a
consequence of $m\sim r^{3}$ and not of regularity as such. The quantum
Oppenheimer--Snyder metric of \cref{ssec:qos} illustrates this point, since it
avoids a big bang without a de~Sitter core and the routes then disagree
qualitatively rather than quantitatively: route N gives a bounce, route A/B sees a
divergent $\Leff$ and no core at all. Prescription dependence is therefore mild in
our gallery precisely because every metric in it shares the same core, and the
$\Delta w_{0}\simeq0.17$ of \cref{sec:routeCimpl} is the mild version of a much larger effect.

The six metrics fall into three groups. The \emph{rational de~Sitter cores}
(Hayward, Bardeen) are analytic or special-function across all routes, reach the
DESI $w_{0}$ band, and behave uniformly. The \emph{Gaussian de~Sitter cores}
(Dymnikova, NSS) regularise the dust cosmology with a finite core but force a
numerical time integral, decay faster than any power, and sit close to $\Lambda$CDM. The two are observationally degenerate. The \emph{special cases} are
Bonanno--Reuter, whose self-critical factor regularises the radiation as well as
the dust cosmology (a property it shares with Hayward and Dymnikova) and whose regularisation
descends from the asymptotic-safety flow rather than from a chosen profile, and
 Jusufi--Singleton, Bardeen-class in every route and the only member of the set
whose Misner--Sharp density can develop a shell, through its self-energy term, hence
the only metric able in principle to cross the phantom divide, though only for
$M/\ell>15$, far outside the core-branch values of \cref{tab:cpl} and in the regime in which Jusufi and Singleton themselves bound the self-energy term, so that with their own prescription no member of the gallery develops a shell.

Several questions remain open and we leave them for future work. The inflationary reading of
route N calls for a perturbation spectrum and for an exit to radiation, both of which require the
non-polynomial theory of \cref{sec:covariant} beyond its homogeneous sector. On the late-time side,
the fiducial scales should be replaced by a fit of $\ell$ to the DESI distances, and the density-shell
geometries singled out by the crossing criterion deserve a study of their own.

Across routes and models, two facts are robust.  The de~Sitter core is universal as a property of the source, every route reads the
same $\Lcore$ from the geometry, and it becomes a primordial acceleration in the one
route where the regular black hole is the whole content. That inflationary phase is the
prediction of route N, and the covariant origin of route N (\cref{sec:covariant}) is
what gives it weight.
And the dynamical dark energy favoured by DESI is, by the crossing criterion, not what a
regular core produces: a monotone core (the generic weak-energy-condition case realised by the regular metrics of~\cite{CasadioKO2025,Easson2026,Firouzjahi2026}) gives thawing quintessence, and a phantom crossing requires a density shell, i.e.\  a hairy rather than a regular black hole. Set against the recent obstruction and
matching results, which establish what a minimal regular-black-hole daughter
cannot do, the present work supplies the complementary positive content: the analytic and cosmographic description of the minimal (route N) model across six
metrics, the non-minimal routes that realise the additional structure the
obstruction requires, and a barrier on the equation of state that none of those analyses
contains. Regular-black-hole cosmology appears, on this evidence, to be first of all a
theory of the early universe. The same length which removes the central singularity
sets the inflationary scale, and what remains at late times is a quintessence which is
compatible with the present data without being singled out by them. In the language
of the obstruction, monotone cores keep the averaged null energy condition and lose
geodesic completeness. Should the data eventually insist on a phantom crossing, it
will have to come from a density shell. We find this a rather clear-cut situation, and a testable one.

\subsection*{Acknowledgments}
S.B.M.\ received support from Vicerrectoría de Investigaciones (VRI) at Pontificia Universidad Javeriana (PUJ) through the VRI06-2024 grant. 

\subsection*{Data availability statement}
No new data were created or analysed in this study. The scripts that reproduce the figures and
tables are available from the authors upon reasonable request.

\end{document}